\documentclass[11pt,reqno]{amsart}
\usepackage{array}
\usepackage{amsmath}
\usepackage{amssymb}
\usepackage{xspace}
\usepackage{graphicx}
\usepackage{psfrag}
\usepackage{longtable}
\usepackage{pstricks,pst-plot,pst-key}
\usepackage[ruled]{algorithm}
\usepackage{algorithmic}
\usepackage{textcomp}
\usepackage{latexsym}

\begin{document}

\bibliographystyle{amsalpha}

\newtheorem{theorem}{Theorem}
\newtheorem{lemma}[theorem]{Lemma}
\newtheorem{claim}[theorem]{Claim}
\newtheorem{conjecture}[theorem]{Conjecture}
\newtheorem{remark}[theorem]{Remark}
\newtheorem{proposition}[theorem]{Proposition}
\newtheorem{property}[theorem]{Property}
\newtheorem{corollary}[theorem]{Corollary}
\newtheorem{definition}[theorem]{Definition}
\newtheorem{problem}[theorem]{Problem}

\newcommand{\oneplus}{\relax}

\newcommand{\muf}{\textsc{MUF}\xspace}
\newcommand{\mufs}{{\muf}s\xspace}
\newcommand{\ra}{\rightarrow}
\newcommand{\fcore}{\ensuremath{F_{\operatorname{c}}}}
\newcommand{\num}{\#}
\newcommand{\aut}{\operatorname{aut}}
\newcommand{\auts}{|\aut H|}
\renewcommand{\a}{\alpha}
\renewcommand{\b}{\beta}
\newcommand{\D}{\Delta}
\newcommand{\E}{\mathbb{E}}
\newcommand{\N}{\mathbb{N}}
\renewcommand{\Re}{\mathbb{R}}
\newcommand{\isom}{\equiv}
\newcommand{\cond}{\; | \;}
\newcommand{\pois}{\operatorname{Po}}
\newcommand{\T}{{\mathcal T}}

\newcommand{\prob}{\mathbb P}
\newcommand{\poly}{\operatorname{poly}}
\def\l{\lambda}
\newcommand{\cut}{Cut\xspace}
\newcommand{\dicut}{Dicut\xspace}
\newcommand{\mmax}{Max\xspace}
\newcommand{\csp}{CSP\xspace}
\newcommand{\maxcut}{\mmax \cut}
\newcommand{\maxdicut}{\mmax \dicut}
\newcommand{\maxcsp}{\mmax \csp}
\newcommand{\maxtcsp}{\mmax 2-\csp}
\newcommand{\maxtsat}{\mmax 2-\sat}
\newcommand{\maxktcsp}{\mmax $(k,2)$-\csp}
\newcommand{\maxkcsp}{\maxktcsp}
\newcommand{\maxttcsp}{\mmax $(2,2)$-\csp}
\newcommand{\maxthreecut}{\mmax 3-\cut}
\newcommand{\sat}{\textsc{sat}\xspace}
\newcommand{\expfn}[1]{\exp\left({#1}\right)}
\newcommand{\ignore}[1]{}
\newcommand{\conference}[1]{}
\newcommand{\fullpaper}[1]{}
\renewcommand{\k}{\ensuremath{\kappa}}
\newcommand{\g}{\ensuremath{\gamma}}
\newcommand{\Be}{\operatorname{Be}}
\newcommand{\G}{{\mathcal G}}
\newcommand{\I}{\ensuremath{\mathcal{M}}}

\newcommand{\ltime}[1]{\ensuremath{O(m+n) \, 2^{m/{#1}}}}
\newcommand{\ktime}[1]{\ensuremath{O(n) \, k^{m/{#1}}}}
\newcommand{\ttime}[1]{\ensuremath{O(n) \, 2^{m/{#1}}}}
\newcommand{\ptime}[1]{\ensuremath{\poly(m+n) \, 2^{m/{#1}}}}

\newcommand{\ie}{\textit{i.e.}}
\newcommand{\mc}{Max 2-CSP\xspace}
\newcommand{\mcr}{Max $(r,2)$-CSP\xspace}
\newcommand{\mcd}{Max $(2,2)$-CSP\xspace}
\newcommand{\eqdef}{:=}
\newcommand{\tmbf}[1]{\textbf{\boldmath{#1}}}
\renewcommand{\O}[1]{O\left(#1\right)}
\newcommand{\Ostar}[1]{{O^\star}(#1)}
\newcommand{\Ostarb}[1]{{O^\star}\big(#1\big)}
\newcommand{\Generalized}{Polynomial\xspace} 

\newcommand{\nvec}{\vec{n}}
\newcommand{\vvec}{\mathbf{v}}
\newcommand{\evec}{\vec{e}}
\newcommand{\yvec}{\vec{y}}
\newcommand{\df}{\vec{d_4}}
\newcommand{\dt}{\vec{d_3}}
\newcommand{\db}{\vec{d_2}}
\newcommand{\dm}{\vec{d_1}}
\newcommand{\steps}{\vec{s}}

\newcommand{\inst}{\ensuremath{\mathcal{M}}\xspace}
\newcommand{\instp}{\ensuremath{\mathcal{M}'}\xspace}
\newcommand{\vstar}{\ensuremath{v^\star}}
\newcommand{\timea}{\ensuremath{O(\oneplus n r^{3+m/5})}\xspace}
\newcommand{\timeb}{\ensuremath{O(\oneplus n r^{5+19m/100})}\xspace}
\newcommand{\ip}{{i+1}}
\newcommand{\im}{{i-1}}
\newcommand{\maxx}{{\operatorname{max}}}
\newcommand{\success}{\ensuremath{\operatorname{success}}\xspace}
\newcommand{\failure}{\ensuremath{\operatorname{failure}}\xspace}

\newcommand{\monad}[3]{{#3}_{#1}({#2})}
\newcommand{\s}[2]{\monad{#1}{#2}{s}}
\renewcommand{\sp}[2]{\monad{#1}{#2}{s'}}
\newcommand{\dyad}[5]{{#5}_{#1 #2}({#3 #4})}
\renewcommand{\ss}[4]{\dyad{#1}{#2}{#3}{#4}{s}}
\newcommand{\ssp}[4]{\dyad{#1}{#2}{#3}{#4}{s'}}
\newcommand{\snought}{s_\emptyset}
\newcommand{\aved}{d} 

\newcommand{\ignorex}[2]{}
\floatname{algorithm}{\ignorex}
\newcommand{\alga}{Algorithm~A\xspace}
\newcommand{\algaa}{Algorithm~A.1\xspace}
\newcommand{\algab}{Algorithm~A.2\xspace}
\newcommand{\algac}{Algorithm~A.3\xspace}
\newcommand{\algb}{Algorithm~B\xspace}
\newcommand{\algba}{Algorithm~B.1\xspace}
\newcommand{\algbb}{Algorithm~B.2\xspace}
\newcommand{\algbc}{Algorithm~B.3\xspace}
\newcommand{\bb}{\ensuremath{\operatorname{B.\!2}}\xspace}
\newcommand{\AB}{\ensuremath{\operatorname{A.\!3}}\xspace}
\newcommand{\half}{\text{\textonehalf}}
\newcommand{\halfe}{\half e}

\newcommand{\ntime}{\ensuremath{\Ostar{r^{19m/100}}}\xspace}
\newcommand{\otime}{\ensuremath{\Ostar{r^{m/5}}}\xspace}
\newcommand{\set}[1]{\{ {#1} \}}
\newcommand{\tw}{\operatorname{tw}}
\newcommand{\oo}{o(1)}
\newcommand{\LPfour}{LP$_4$}
\newcommand{\xstar}{x^*}
\newcommand{\Ostarr}[2]{\ensuremath{\Ostar{ {#1}^{#2}}}}

\newcounter{rownum}
\setcounter{rownum}{0}
\newcommand{\rownum}{\addtocounter{rownum}{1} \arabic{rownum}}

\long\def\tabrownonumx #1 #2 #3 #4 #5 #6 %
{ \tabrowl #1 #2 #3 #4 #5 #6 \tabrowr  }

\long\def\tabrownonum #1 #2 #3 #4 #5 #6 %
{ & \tabrowl #1 #2 #3 #4 #5 #6 \tabrowr  }

\long\def\tabrow #1 #2 #3 #4 #5 #6 %
{\rownum & \tabrowl #1 #2 #3 #4 #5 #6 \tabrowr  }

\long\def\tabrowl #1 #2 #3 #4 #5 #6%
{$#1$ & $#2$ & $#3$ & $#4$ & $#5$ & $#6$ &}

\long\def\tabrowr #1 #2 #3 #4 #5 #6 #7%
{$#1$ & $#2$ & $#3$ & $#4$ & $#5$ & $#6$ & $#7$ \\}

\long\def\boldtabrow #1 #2 #3 #4 #5 #6 %
{\rownum & \boldtabrowl #1 #2 #3 #4 #5 #6 \boldtabrowr  }

\long\def\boldtabrowl #1 #2 #3 #4 #5 #6%
{$\mathbf{#1}$ & $\mathbf{#2}$ & $\mathbf{#3}$ & $\mathbf{#4}$
  & $\mathbf{#5}$ & $\mathbf{#6}$ &}

\long\def\boldtabrowr #1 #2 #3 #4 #5 #6 #7%
{$\mathbf{#1}$ & $\mathbf{#2}$ & $\mathbf{#3}$ & $\mathbf{#4}$
  & $\mathbf{#5}$ & $\mathbf{#6}$& $\mathbf{#7}$  \\}

\title%
[LP design of fast Max 2-CSP algorithms]
{Linear-programming design and analysis of 
\\
fast algorithms for Max 2-CSP}


\author[Alexander D. Scott]{Alexander D. Scott$^\star$}
\thanks{$^\star$ Research supported in part by EPSRC grant GR/S26323/01}%
\address[Alexander D. Scott]{
Mathematical Institute\\
University of Oxford\\
24-29 St Giles'\\ 
Oxford OX1 3LB, UK}
\email{scott@maths.ox.ac.uk}

\author[Gregory B. Sorkin]{Gregory B. Sorkin%
}
\address[Gregory B. Sorkin]{
Department of Mathematical Sciences \\
IBM T.J.\ Watson Research Center \\
Yorktown Heights NY 10598, USA}
\email{sorkin@watson.ibm.com}

\thispagestyle{empty}

\begin{abstract}
The class \mcr,
or simply \mc,
consists of constraint satisfaction problems 
with at most two $r$-valued variables per clause.
For instances with $n$ variables and $m$ binary clauses, 
we present an \timeb-time algorithm
which is the fastest polynomial-space algorithm 
for many problems in the class, including Max Cut.
The method also proves a treewidth bound 
$\tw(G) \leq (13/75+\oo)m$,
which gives a faster \mc algorithm that uses exponential space:
running in time $\Ostar{2^{(13/75+\oo)m}}$, 
this is fastest for most problems in \mc.
Parametrizing in terms of $n$ rather than~$m$, 
for graphs of average degree $\aved$ we show a simple algorithm running time 
$O^\star\big(2^{\left(1-\frac{2}{\aved+1}\right)n}\big)$,
the fastest polynomial-space algorithm known.

In combination with ``\Generalized CSP{s}'' introduced in a 
companion paper, these algorithms also allow
(with an additional polynomial-factor overhead in space and time)
counting and sampling, 
and the solution of problems like Max Bisection that 
escape the usual CSP framework.

Linear programming is key to the design 
as well as the analysis of the algorithms.
\end{abstract}

\maketitle

\section{Introduction} 

A recent line of research has been to speed up exponential-time algorithms
for sparse instances of 
maximization problems such as Max 2-Sat and Max Cut.
The typical method is to 
repeatedly \emph{transform} an instance to a smaller one
or \emph{split} it into several smaller ones
(whence the exponential running time)
until trivial instances are reached;
the reductions are then reversed to recover a solution 
to the original instance.
In \cite{random03} we introduced a new such method, 
distinguished by the fact that reducing an instance of Max Cut, 
for example, results in a problem that may not belong to Max Cut, 
but where the reductions are closed over a larger class,
\mc, of constraint satisfaction problems with at most two variables per clause.
This allowed the reductions to be simpler, fewer, and more powerful. 
The algorithm ran in time $\Ostar{2^{m/5}}$ 
(time $\Ostar{r^{m/5}}$ for $r$-valued problems), 
making it the fastest for Max Cut, and tied 
(at the time) for Max 2-Sat.

In this paper we present a variety of results
on faster exponential-time CSP algorithms
and on treewidth.  Our approach uses linear programming in both the design and
the analysis of the algorithms.

\subsection{Results}
The running times for our algorithms 
depend on the space allowed, and are
summarized in Table~\ref{resultstable}.
The $\Ostar{\cdot}$ notation,
which ignores leading polynomial factors,
is defined in Section~\ref{notation}.

\newcommand{\strutt}[1]{\rule{0em}{#1}}
\newcommand{\padhline}{\rule{0em}{1.2em}}
\newcommand{\padtableend}{\vspace{0.3em}}

\begin{table}
\begin{tabular}{|r||l|ll||l|l|}
\hline
\multicolumn{6}{|c|}{\strutt{1.1em}edge parametrized ($m$)} \\
\hline \hline \padhline
\strutt{1.1em}
problem & time (exact) & \multicolumn{2}{c||}{time (numerical)} & space 
 & reference \\
\hline \hline \padhline
\mcr       & \Ostarr r {19m/100} 
 & \Ostarr r {0.19m} & \Ostarr r {m/5.2631} & linear 
 & Theorem~\ref{maintheorem} \\
$\Delta \leq 4$ & \Ostarr r {3m/16} 
 & \Ostarr r {0.1875m} & \Ostarr r {m/5.3333} & & \\
$\Delta \leq 3$ & \Ostarr r {m/6} 
 & \Ostarr r {0.1677m} & \Ostarr r {m/6} & & \\
\hline \hline \padhline
\mcr      & \Ostarr r {(13/75+\oo)m} 
 & \Ostarr r {0.1734m} & \Ostarr r {m/5.7692} & exponential 
 & Corollary~\ref{generic} \\
$\Delta \leq 4$ & \Ostarr r {(1/6+\oo)m} 
 & \Ostarr r {0.1667m} & \Ostarr r {m/5.9999} & & \\
$\Delta \leq 3$ & \Ostarr r {(1/9+\oo)m} 
 & \Ostarr r {0.1112m} & \Ostarr r {m/8.9999} & & \\
\hline 
\multicolumn{6}{c}{\mbox{ }}\\
\hline
\multicolumn{6}{|c|}{\strutt{1.1em}vertex parametrized ($n$)} \\
\hline \hline \padhline
\strutt{1.5em}
\mcr 
 & $\Ostarb{r^{ (1-\frac2{\aved+1}) n}}$
 &  &  & polynomial 
 & Theorem~\ref{n-param} \\
\strutt{1.2em}
 & $\Ostarb{r^{(d-2)n/4}}$
 &  &  & polynomial 
 & \cite{random03,linear} \\
\hline
\end{tabular}
\padtableend
\caption{
Exact bounds and numerical bounds (in two forms)
on the running times of our \mcr algorithms.
All of these are the best known.
Throughout this paper,
$m$ denotes the number of 2-clauses and $n$ the number of variables;
$\Delta$ denotes the maximum number of 2-clauses on any variable,
and $\aved$ the average number.
}
\label{resultstable}
\end{table}

For \mc
we give an $\Ostar{r^{19m/100}}$-time, linear-space algorithm.
This is the fastest pol\-y\-no\-mi\-al-space algorithm known for 
Max Cut, Max Dicut, Max 2-Lin,
less common problems such as Max Ones 2-Sat,
weighted versions of all these, 
and of course general \mc;
more efficient algorithms are known for only a few problems,
such as Maximum Independent Set
and Max 2-Sat.  
If exponential space is allowed, 
we give an algorithm running in time $\Ostar{r^{(13/75+\oo)m}}$
and space $\Ostar{r^{(1/9+\oo)m}}$;
it is the fastest exponential-space algorithm
known for most problems in \mc 
(including those listed above for the polynomial-space algorithm).

These bounds have connections with treewidth, 
and we prove that the treewidth of an $m$-edge graph $G$ 
satisfies
$\tw(G) \leq 3+19m/100$
and
$\tw(G) \leq (13/75+\oo)m$.
(The second bound is clearly better for large~$m$.)

For both treewidth and algorithm running time
we provide slightly better results for
graphs of maximum degree $\D(G)=3$ and $\D(G)=4$.

In combination with a ``\Generalized CSP'' approach presented in 
a companion paper~\cite{CountingArxiv,CountingArxiv2},
the algorithms here also enable 
(with an additional polynomial-factor overhead in space and time)
counting CSP solutions of each possible cost; 
sampling uniformly from optimal solutions; 
sampling from all solutions according to the Gibbs measure
or other distributions; 
and solving problems that do not fall into the \mc framework, 
like Max Bisection, Sparsest Cut, judicious partitioning, 
Max Clique (without blowing up the input size),
and multi-objective problems.  We refer to \cite{CountingArxiv,CountingArxiv2}
for further details.

Our emphasis is on running time 
parametrized in terms of the number of edges~$m$, 
but we also have results for 
parametrization in terms of the number of edges~$n$
(obtained largely independently of the methods in the rest of the paper).
The main new result is a \mc algorithm running in time
$\Ostarb{r^{ (1-\frac2{\aved+1}) n}}$ (Theorem~\ref{n-param}),
where $d$ is the average number of appearances of each variable
in 2-clauses. 
Coupled with an older algorithm of ours (see \cite{random03,linear}) 
with running time $\Ostarb{r^{ (1-\frac2{d+2}) n}}$, 
this is the best known polynomial-space algorithm.

\subsection{Techniques}
We focus throughout on the ``constraint graph'' supporting a CSP instance.
Our algorithms use several simple transformation rules, 
and a single splitting rule.  
The transformation rules replace an instance by an equivalent instance 
with fewer variables; 
our splitting rule produces several instances, 
each with the same, smaller, constraint graph.
In a simple recursive CSP algorithm, then,
the size of the CSP ``recursion tree'' is exponential in the
number of splitting reductions on the graph.
The key step in the analysis of our {earlier $\otime$ algorithm}
was to show that the number of splitting reductions for 
an $m$-edge graph can be no more than $m/5$.

We used a linear programming (LP) analysis to derive an upper bound
on how large the number of splitting reductions can be.
Each reduction affects the degree sequence of the graph in a simple way, 
and the fact that the number of vertices of each degree 
is originally non-negative and finally 0 is enough to derive the
$m/5$ bound.

It is not possible to improve upon the $m/5$ bound on the 
\emph{number} of splitting reductions, since there are examples
achieving the bound. 
However, we are able to obtain a smaller bound on the reduction ``depth'' 
(described later), and
the running time of a more sophisticated algorithm is exponential
in this depth.
Analysis of the reduction depth must take into account the component structure 
of the CSP's constraint graph.  
The component structure is not naturally captured by the LP analysis,
which considers the (indivisible) degree sequence of the full graph
(the usual argument that in case of component division 
``we are done, by induction'' cannot be applied)
but a slight modification of the argument resolves the difficulty.

We note that
the LP was essential in the \emph{design} of the new algorithm as well 
as its analysis.
The support of the LP's \emph{primal} 
solution indicates the particular reductions 
that contribute to the worst case.
With a ``bad'' reduction identified, we do two things in parallel: 
exclude the reduction from the LP to see if an improved 
bound would result, 
and apply some actual thinking to see if it is possible to avoid
the bad reduction. 
Since thinking is difficult and time-consuming, it is nice that the
LP can be modified and re-run in a second to determine whether any gain would 
actually result. 
Furthermore,
the LP's \emph{dual} solution gives an (optimal) set of weights, 
for edges and for vertices of each degree,
for a ``Lyapunov'' or ``potential function'' 
proof of the depth bound.
The potential-function proof method is well established (in the
exponential-time algorithm context the survey \cite{FominSurvey} calls it
a ``measure and conquer'' strategy), and the LP method gives an 
efficient and provably optimal way of carrying it out.

The LP method presented 
is certainly applicable to reductions 
other than our own, and we also hope to see it applied
to algorithm design and analysis in contexts other than 
exponential-time algorithms and CSPs.  
(For a different use of LPs in automating extremal constructions,
see~\cite{SSTW00}.)

\subsection{Literature survey} \label{literature}

We are not sure where the class $(a,b)$-CSP was first introduced,
but this model, where each variable has at most $a$ possible colors
and there are general constraints each involving at most $b$ variables,
is extensively exploited for example in Beigel and Eppstein's
$\Ostar{1.3829^n}$-time 3-coloring algorithm~\cite{Epps}.
Finding relatively fast exponential-time algorithms for NP-hard problems
is a field of endeavor that includes
Sch\"oning's famous randomized algorithm for 3-Sat~\cite{schoening},
taking time $\Ostar{(4/3)^n}$ for an instance on $n$ variables.

Narrowing the scope to \mc{s} with time parametrized in~$m$,
we begin our history with an algorithm of 
Niedermeier and Rossmanith~\cite{NiRo00}:
designed for Max Sat generally,
it solves Max 2-Sat instances in time $\Ostar{2^{0.348 m}}$.
The Max 2-Sat result was improved by Hirsch 
to $\Ostar{2^{m/4}}$~\cite{Hirsch}.
Gramm, Hirsch, Niedermeier and Rossmanith
showed how to solve Max 2-Sat in time $\Ostar{2^{m/5}}$,
and used a transformation from
Max Cut into Max 2-Sat to allow Max Cut's solution
in time $\Ostar{2^{m/3}}$~\cite{Gramm03}.
Kulikov and Fedin showed how to solve Max Cut in time 
$\Ostar{2^{m/4}}$~\cite{fedin}. 
Our own \cite{random03} improved the Max Cut time 
(and any Max 2-CSP) to $\Ostar{2^{m/5}}$.
Kojevnikov and Kulikov recently improved the Max 2-Sat time to 
$\Ostar{2^{m/5.5}}$ \cite{KK06}; 
at the time of writing this is the fastest.

We now improve the time for Max Cut to $\Ostar{2^{19m/100}}$.
We also give linear-space algorithms for all of \mc running in 
time $\ntime$,
as well as faster but exponential-space algorithms running in
time $\Ostar{2^{(13/75+\oo)m}}$. 
and space $\Ostar{2^{(1/9+\oo)m}}$. 
All these new results are the best currently known.

A technical report of Kneis and Rossmanith \cite{Kneis05} 
(published just months after our~\cite{FasterIBM}),
and a subsequent paper 
of Kneis, M\"olle, Richter and Rossmanith \cite{Kneis4},
give results overlapping with those in \cite{FasterIBM}
and the present paper.
They give algorithms applying to several problems in \mc,
with claimed running times of $\Ostar{2^{19m/100}}$
and (in exponential space) $\Ostar{2^{13m/75}}$.
The papers are widely cited but confuse the literature to a degree.
First, the authors were evidently unaware of \cite{FasterIBM}.
\cite{Kneis05} cites our much earlier conference paper~\cite{random03}
(which introduced many of the ideas extended in \cite{FasterIBM}
and the present paper)
but overlooks both its $\Ostar{2^{m/5}}$ algorithm and its II-reduction
(which would have extended their results to all of \mc). 
These oversights are repeated in \cite{Kneis4}.
Also, both papers have a reparable but fairly serious flaw,
as they overlook the ``component-splitting'' case~C4
of Section~\ref{breduct} 
(see Section~\ref{secb} below). 
Rectifying this means adding the missing case, 
modifying the algorithm to work component-wise, 
and analyzing ``III-reduction depth'' rather than the total number 
of III-reductions 
--- the issues that occupy us throughout Section~\ref{secb}.
While treewidth-based algorithms have a substantial history
(surveyed briefly in Section~\ref{sec:treewidth}), 
\cite{Kneis05} and \cite{Kneis4} motivate
our own exploration of treewidth, 
especially Subsection~\ref{cubic}'s
use of Fomin and H{\o}ie's~\cite{Fomin}.

We turn our attention briefly to 
algorithms parametrized in terms of the number $n$ of vertices (or variables), 
along with the average degree~$\aved$
(the average number of appearances of a variable in 2-clauses)
and the maximum degree~$\Delta$.
A recent result of 
Della Croce, Kaminski, and Paschos~\cite{DCKP06}
solves Max Cut (specifically) in time 
$\Ostarb{2^{mn/(m+n)}} = \Ostarb{2^{ (1-\frac2{d+2}) n}}$.
Another recent paper, 
of F\"urer and Kasiviswanathan~\cite{FK07},
gives a running-time bound of 
$\Ostarb{2^{ (1-\frac{1}{\aved-1}) n}}$
for any \mc
(where $\aved>2$ and the constraint graph is connected,
per personal communication).
Both of these results are superseded by 
the \mc algorithm of Theorem~\ref{n-param},
with time bound
$\Ostarb{2^{ (1-\frac2{\aved+1}) n}}$,
coupled with another of our algorithms from \cite{random03,linear},
with running time $\Ostarb{2^{ (1-\frac2{d+2}) n}}$.
A second algorithm from~\cite{DCKP06},
solving Max Cut in time $\Ostarb{2^{ (1-2/\Delta) n}}$,
remains best for ``nearly regular'' instances
where $\Delta \leq d+1$.

Particular problems within \mc can often be solved faster.
For example, an easy tailoring of our $\Ostar{r^{19m/100}}$
algorithm to weighted Maximum Independent Set
runs in time $\Ostar{2^{3n/8}}$
(see Corollary~\ref{MIS}), which is $\Ostar{1.2969^n}$. 
This improves upon an older algorithm of
Dahll\"of and Jonsson~\cite{Dahllof2002},
but is not as good as 
the $\Ostar{1.2561^n}$ algorithm of 
Dahll\"of, Jonsson and Wahlstr{\"o}m~\cite{Jonsson05}
or the $\Ostar{1.2461^n}$ algorithm of 
F\"urer and Kasiviswanathan~\cite{Furer05}.
(Even faster algorithms are known for unweighted~MIS.)

The elegant algorithm of Williams~\cite{Williams}, like our algorithms,
applies to all of \mc.
It is the only known algorithm to treat \emph{dense} instances of 
\mc relatively efficiently, and also enjoys some of the strengths of our 
\Generalized CSP extension~\cite{CountingArxiv,CountingArxiv2}.
It intrinsically requires exponential space, 
of order ${2^{2n/3}}$, and 
runs in time $\Ostar{2^{\omega n/3}}$, where $\omega<2.376$ is the
matrix-multiplication exponent. 
Noting the dependency on $n$ rather than $m$, 
this algorithm is faster than our polynomial-space algorithm
if the average degree is above $2 (\omega/3)/(19/100)) < 8.337$,
and faster than our exponential-space algorithm 
if the average degree is above $9.139$. 

An early version of our results 
was given in the technical report~\cite{FasterIBM},
and a conference version 
appeared as~\cite{FasterESA}.

\subsection{Outline}
In the next section we define the class \mc, 
and in Section~\ref{reductionsec} we introduce the reductions our algorithms 
will use.
In Section~\ref{seca} we define and analyze the \timea algorithm 
of \cite{random03}
as a relatively gentle introduction to the tools, including the LP analysis. 
The \timeb algorithm is presented 
in Section~\ref{secb}; 
it entails a new focus on components of the constraint graph, 
affecting the algorithm and the analysis.
Section~\ref{n-sec} digresses to consider algorithms with run time
parametrized by the number of vertices rather than edges; 
by this measure, it gives the fastest known polynomial-space algorithm 
for general \mc instances.
Section~\ref{sec:treewidth} presents corollaries pertaining to
the treewidth of a graph
and
the exponential-space $\Ostar{r^{(13/75+\oo)m}}$ algorithm.
Section~\ref{conclusions} recapitulates, and considers the potential 
for extending the approach in various ways.

\section{Max $(r,2)$-CSP} \label{cspsec}

The problem {Max Cut} is to
partition the vertices of a given graph into two classes
so as to maximize the number of edges ``cut'' by the partition.
Think of each \emph{edge} 
as being a \emph{function} on the classes (or ``colors'') of its endpoints,
with value 1 if the endpoints are of different colors, 0 if they are the same:
Max Cut is equivalent to finding a 2-coloring of the vertices which maximizes
the sum of these edge functions.
This view naturally suggests a generalization.

An \emph{instance} $(G,S)$ of {\mcr} 
is given by a ``constraint'' graph $G=(V,E)$
and a set $S$ of ``score'' functions.
Writing $[r]=\{1,\ldots,r\}$ for the set of available colors,
we have a ``dyadic'' score function $s_e: [r]^2 \ra \Re$
for each edge $e \in E$,
a ``monadic'' score function $s_v: [r] \ra \Re$
for each vertex $v \in V$,
and finally a single ``niladic'' score ``function'' $\snought: [r]^0 \ra \Re$
which takes no arguments and is just a constant convenient for bookkeeping.

A \emph{candidate solution} is a function $\phi: V \ra [r]$
assigning ``colors'' to the vertices
(we call $\phi$ an ``assignment'' or ``coloring''),
and its \emph{score} is
\begin{align}
s(\phi) :=
 \snought + \sum_{v \in V} s_v(\phi(v)) 
  + \sum_{uv \in E} s_{uv}(\phi(u),\phi(v)) .
\end{align}
An \emph{optimal solution} $\phi$ is one which maximizes $s(\phi)$.

We don't want to belabor the notation for edges, but
we wish to take each edge just once, and
(since $s_{uv}$ need not be a symmetric function)
with a fixed notion of which endpoint is ``$u$'' and which is ``$v$''.
We will typically assume that $V=[n]$ and any edge $uv$
is really an ordered pair $(u,v)$ with $1\leq u<v \leq n$; we will also feel free to
abbreviate $s_{uv}(C,D)$ as $s_{uv}(CD)$, etc.

Henceforth we will simply write \mc for the class \mcr.
The ``2'' here 
refers to score functions'
taking 2 or fewer arguments: 3-Sat, for example, is out of scope.
Replacing 2 by a larger value would mean replacing the constraint
graph with a hypergraph, and changes the picture significantly.

An obvious computational-complexity issue 
is raised by our allowing scores to be arbitrary
\emph{real} values.
Our algorithms will add, subtract, and compare these scores,
never introducing a number larger in absolute value than the
sum of the absolute values of all input values,
and we assume that each such operation can be done in time 
and space~$O(1)$.
If desired, scores may be limited to integers,
and the length of the integers factored in to the algorithm's complexity,
but this seems uninteresting and we will not remark on it further.

\subsection{Notation}
\label{notation}
We reserve the symbols
$G$ for the constraint graph of a \mc instance,
$n$ and $m$ for its numbers of vertices and edges,
$[r]=\{1,\ldots,r\}$ for the allowed colors of each vertex,
and $L=1+nr+m r^2$ for the input length.
Since a CSP instance with $r<2$ is trivial, 
we will assume $r \geq 2$ as part of the definition. 

For brevity, we will often write ``$d$-vertex'' 
in lieu of ``vertex of degree~$d$''.
We write $\D(G)$ for the maximum degree of~$G$.

The notation $\Ostar{\cdot}$ suppresses polynomial factors in any parameters,
so for example $\Ostar{r^{cn}}$ may mean $O(r^3 n \: r^{cn})$.
To avoid any ambiguity in multivariate $O(\cdot)$ expressions, 
we take a strong interpretation that
that $f(\cdot)=O(g(\cdot))$ if there exists some constant $C$ 
such that $f(\cdot) \leq C g(\cdot)$ 
for \emph{all} values of their (common) arguments.
(To avoid some notational awkwardness, 
we disallow the case $n=0$, but allow~$m=0$.)

\subsection{Remarks}

Our assumption of an undirected constraint graph is sound
even for a problem such as {Max Dicut} (maximum directed cut).
For example, for Max Dicut a directed edge $(u,v)$ with $u<v$ 
would be expressed by the score function $s_{uv}(\phi(u),\phi(v))=1$ 
if $(\phi(u),\phi(v))=(0,1)$ and $s_{uv}(\phi(u),\phi(v))=0$ otherwise;
symmetrically, a directed edge $(v,u)$, again with $u<v$, 
would have score $s_{uv}(\phi(u),\phi(v))=1$ 
if $(\phi(u),\phi(v))=(1,0)$ and score 0 otherwise.

There is 
no loss of generality in assuming that an input instance
has a simple constraint graph (no loops or multiple edges),
or by considering only maximization and not minimization problems.

\newcommand{\calf}{\ensuremath{\mathcal{F}}}
\newcommand{\fsat}{\calf-Sat\xspace}
\newcommand{\maxfsat}{\calf-Max-Sat\xspace}
\newcommand{\minfsat}{\calf-Min-Sat\xspace}
\newcommand{\fmaxones}{\calf-Max-Ones\xspace}
\newcommand{\fminones}{\calf-Min-Ones\xspace}

Readers familiar with the class \fsat
(see for example Marx~\cite{marx}, Creignou~\cite{creignou}, 
or Khanna~\cite{khanna})
will realize that when the arity of \calf\ is limited to~2,
\mc contains \fsat, \maxfsat and \minfsat; 
this includes Max 2-Sat and Max 2-Lin
(satisfying as many as possible of $m$
2-variable linear equalities and/or inequalities).
\mc also contains \fmaxones; for example Max-Ones-2-Sat.
Additionally, \mc contains similar problems where we
maximize the weight rather than merely the number of satisfied clauses.

The class \mc is surprisingly flexible, 
and in addition to Max Cut and Max 2-Sat includes problems 
like MIS and minimum vertex cover
that are not at first inspection structured around pairwise constraints.
For instance, to model MIS as a \mc, 
let $\phi(v)=1$ if vertex $v$ is to be included in the independent set,
and 0 otherwise;
define vertex scores $s_v(\phi(v)) = \phi(v)$;
and define edge scores $s_{uv}(\phi(u),\phi(v)) = -2$ if $\phi(u)=\phi(v)=1$,
and 0 otherwise.

\section{Reductions} \label{reductionsec}
As with most of the works surveyed above,
our algorithms are based on progressively reducing an instance
to one with fewer vertices and edges until the instance becomes trivial.
Because we work in the general class \mc
rather than trying to stay within a smaller class such as Max 2-Sat 
or Max $k$-Cut,
our reductions are simpler and fewer than is typical.
For example, \cite{Gramm03} uses seven reduction rules;
we have just three
(plus a trivial ``0-reduction'' that other works may treat implicitly).
The first two reductions each produce equivalent instances with one vertex
fewer, 
while the third produces a set of $r$ instances,
each with one vertex fewer, 
some one of which is equivalent to the original instance.
We expand the previous notation $(G,S)$ for an instance to
$(V,E,S)$, where $G=(V,E)$.

\begin{description}

\item[Reduction 0 (transformation)]
This is a trivial ``pseudo-reduction''.
If a vertex $y$ has degree~0 (so it has no dyadic constraints),
then set $\snought = \snought + \max_{C \in [r]} \s yC$
and delete $y$ from the instance entirely.
\item[Reduction I]
Let $y$ be a vertex of degree~1, with neighbor~$x$.
Reducing $(V,E,S)$ on $y$ results in a new problem $(V',E',S')$
with $V' = V \setminus y$ and $E' = E \setminus xy$.
$S'$ is the restriction of $S$ to $V'$ and $E'$, except that
for all colors $C \in [r]$
we set
\begin{align*}
\sp xC & = \s xC + \max_{D \in [r]} \{\ss xyCD + \s yD\} .
\end{align*}

Note that any coloring $\phi'$ of $V'$ can be extended to
a coloring $\phi$ of $V$ in $r$ ways, depending on the color assigned to~$y$.
Writing $(\phi',D)$ for the extension in which $\phi(y)=D$,
the defining property of the reduction is that
$s'(\phi') = \max_D s(\phi',D)$.
In particular,
$\max_{\phi'}s'(\phi') = \max_{\phi}s(\phi)$,
and an optimal coloring $\phi'$ for the instance $(V',E',S')$
can be extended to an optimal coloring $\phi$ for $(V,E,S)$.

\begin{center}
\psfrag{x}[bc][bc]{$x$}
\psfrag{y}[bc][bc]{$y$}
\includegraphics[height=1.0in]{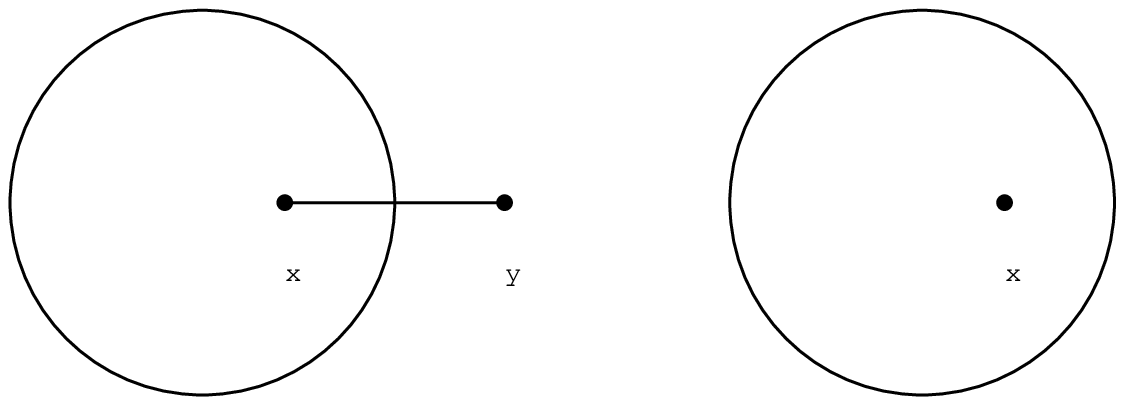}
\end{center}

\item[Reduction II (transformation)]
Let $y$ be a vertex of degree~2, with neighbors~$x$ and~$z$.
Reducing $(V,E,S)$ on $y$ results in a new problem $(V',E',S')$
with $V' = V \setminus y$ and $E' = (E  \setminus \{xy,yz\}) \cup \{xz\}$.
$S'$ is the restriction of $S$ to $V'$ and $E'$, except that
for $C,D \in [r]$ 
we set
\begin{align}
\ssp xzCD & =
 \ss xzCD +
\max_{F \in [r]} \{
\ss xyCF + \ss yzFD + \s yF\}
 \label{IIreduction}
\end{align}
if there was already an edge~$xz$, 
discarding the first term $\ss xzCD$ 
if there was not.

As in Reduction~I,
any coloring $\phi'$ of $V'$ can be extended to $V$ in $r$ ways,
according to the color $F$ assigned to~$y$,
and the defining property of the reduction is that
$s'(\phi') = \max_F s(\phi',F)$.
In particular,
$\max_{\phi'}s'(\phi') = \max_{\phi}s(\phi)$,
and an optimal coloring $\phi'$ for $(V',E',S')$
can be extended to an optimal coloring $\phi$ for $(V,E,S)$.

\vspace*{0.2cm}
\begin{center}
\psfrag{x}[bc][bc]{$x$}
\psfrag{y}[bc][bc]{$y$}
\psfrag{z}[bc][bc]{$z$}
\includegraphics[height=1.0in]{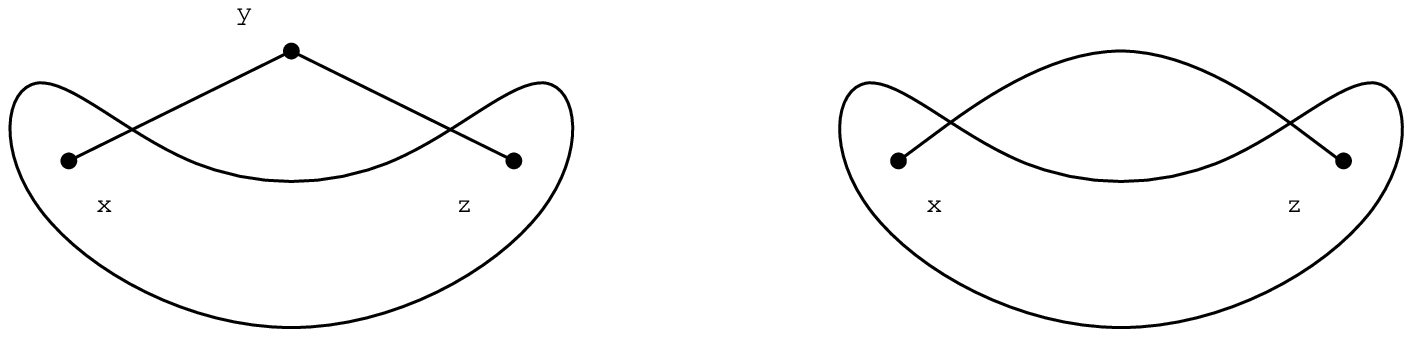}
\end{center}

\newcommand{\sR}[2]{\monad{#1}{#2}{(s^R)}}
\newcommand{\sB}[2]{\monad{#1}{#2}{(s^B)}}

\item[Reduction III (splitting)]
Let $y$ be a vertex of degree 3 \emph{or higher}.
Where reductions I and II each had a single reduction of $(V,E,S)$
to $(V', E', S')$, here
we define $r$ different reductions: for each color $C$ there is a
reduction of $(V,E,S)$ to $(V', E', S^C)$
corresponding to assigning the color $C$ to~$y$.
We define $V' = V \setminus y$, and
$E'$ as the restriction of $E$ to $V \setminus y$.
$S^C$ is the restriction of $S$ to $V \setminus y$,
except that we set
\begin{align*}
(s^C)_0 &= \snought + \s yC ,
\\ \intertext{and, for every neighbor $x$ of $y$ and every $D \in [r]$,}
\monad{x}{D}{(s^C)} &= \s xD + \ss xyDC .
\end{align*}

As in the previous reductions,
any coloring $\phi'$ of $V \setminus y$ can be extended to $V$ in $r$ ways:
for each color $C$ there is an extension
$(\phi',C)$, where color $C$ is given to~$y$.  We then have
(this is different!) $s^C(\phi') = s(\phi',C)$, and
furthermore,
\begin{align*}
\max_C \max_{\phi'} s^C(\phi')
&= \max_{\phi}s(\phi) ,
\end{align*}
where an optimal coloring on the left
\emph{is} an optimal coloring on the right.

\vspace*{0.2cm}
\begin{center}
\psfrag{x}[bc][bc]{$y$}
\psfrag{y}[bc][bc]{$x$}
\psfrag{z}[bc][bc]{$z$}
\psfrag{0}[tl][tl]{} 
\psfrag{1}[tl][tl]{} 
\includegraphics[height=0.90in]{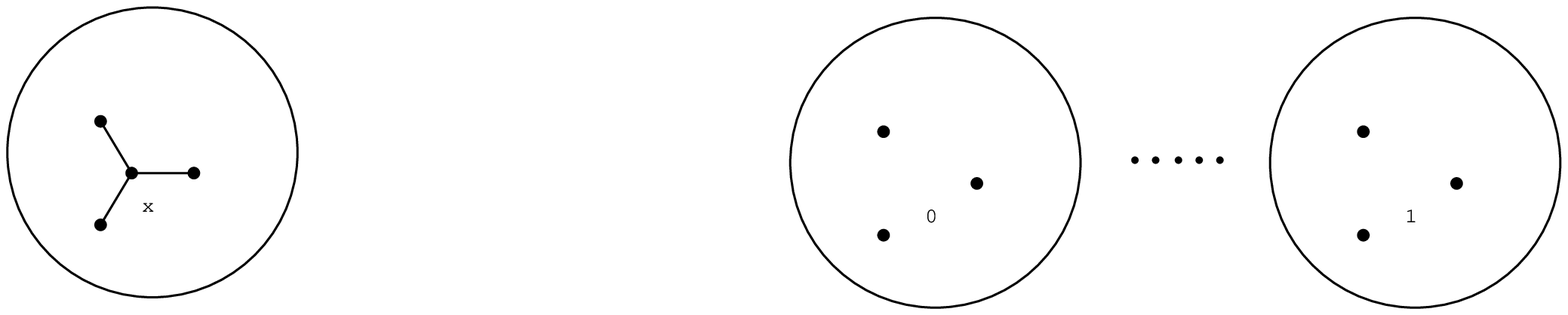}
\\
\hspace*{0.6cm}$(V,E,S)$%
\hspace*{4.5cm}$(V',E',S^1)$%
\hspace*{1.7cm}$(V',E',S^r)$%
\hspace*{0.5cm}
\end{center}

\vspace{0.4cm}
\end{description}

Note that each of the reductions above has a 
well-defined effect on the constraint graph of an instance: 
A 0-reduction deletes its (isolated) vertex; 
a I-reduction deletes its vertex (of degree~1); 
a II-reduction contracts away its vertex (of degree~2);
and a III-reduction deletes its vertex (of degree 3 or more), 
independent of the ``color'' of the reduction.
That is, all the CSP reductions have graph-reduction counterparts 
depending only on the constraint graph and the reduction vertex.

\section{An \timea Algorithm}
\label{seca}

As a warm-up to our \ntime algorithm,
in this section we will present \alga, which will run in time 
\timea and space~$O(L)$. 
(Recall that $L=1+nr+mr^2$ is the input length.)
Roughly speaking, 
a simple recursive algorithm for solving an input instance 
could work as follows.
Begin with the input problem instance.

\pagebreak[2]
Given an instance $\inst = (G,S)$: 
\nopagebreak[4]
\begin{enumerate}
\item
If any reduction of type 0,
I or II is possible (in that order of preference),
apply it to reduce $\inst$ to $\inst'$,
recording certain information about the reduction.
Solve $\inst'$ recursively,
and use the recorded information to 
reverse the reduction and extend the solution to one for~$\inst$.
\item \label{recurstep}
If only a type III reduction is possible,
reduce (in order of preference) 
on a vertex of degree 5 or more, 4, or~3.
For $i \in [r]$,
recursively solve each of the instances $\inst^i$ in turn, 
select the solution with the largest score,
and use the recorded information to
reverse the reduction and extend the solution to one for~$\inst$.
\item \label{laststep}
If no reduction is possible then the graph has no vertices,
there is a unique coloring (the empty coloring),
and the score is $\snought$ (from the niladic score function).
\end{enumerate}

If the recursion depth 
--- the number of III-reductions ---
is $\ell$, the recursive algorithm's running 
time is $\Ostar{r^\ell}$.
Thus in order to prove an \otime bound on running time, 
it is enough to prove that $\ell\le m/5$.  
We prove this bound
in Lemma~\ref{depth5} in Section~\ref{Adepth}.
(The preference order for type III reductions described above is needed to 
obtain the bound.)

In order to obtain our more precise \timea bound on running time, we must be a little more careful
with the description of implementation and data storage.  Thus
Sections \ref{complications} to \ref{secAphase3} 
deal with the additional difficulties arising from running in linear space
and with a small polynomial factor for running time.
A reader willing to take this for granted, 
or who is primarily interested in the 
exponent in the \ntime bound, 
can skip directly to  Section~\ref{Adepth}.

\subsection{Linear space} \label{complications}
If the recursion depth is~$\ell$,
a straightforward recursive implementation would use
greater-than-linear space, namely~$\Theta(\ell L)$.
Instead, when the algorithm has reduced on a vertex~$v$, 
the reduced instance should be the only one maintained, 
while the pre-reduction instance should be reconstructible
from compact ($O(1)$-sized) information stored in the data structure for~$v$.

\subsection{Phases} \label{first phase}
For both efficiency of implementation and ease of analysis,
we define \alga 
as running in three phases.
As noted at the end of Section~\ref{reductionsec}, 
the CSP reductions have graph-reduction counterparts.
In the first phase we merely perform such \emph{graph} reductions.
We reduce on vertices in the order of preference given earlier:
0-reduction (on a vertex of degree~0);
I-reduction (on a vertex of degree~1);
II-reduction (on a vertex of degree~2);
or (still in order of preference)
III-reduction on a vertex of degree 5 or more, 4, or~3.
The output of this phase is simply the sequence of vertices on which we reduced.

The second phase finds the optimal \emph{cost} recursively, 
following the reduction sequence of the first phase; 
if there were $\ell$ III-reductions in the first phase's reduction sequence, 
the second phase runs in time $\Ostar{r^\ell}$.
The third phase is similar to the second phase and returns 
an optimal \emph{coloring}.

\subsection{First phase}
\label{details}

In this subsection we show that
a sequence of reductions following the stated preference order can be
constructed in linear time and space
by \algaa.  (See displayed pseudocode, and details in Claim~\ref{Aphase1}.)

\begin{algorithm*}
\caption{\algaa: \alga, first phase}
\label{A1}
\begin{algorithmic}[1]
\STATE \textbf{Input} a constraint graph~$G$.
\IF{$G$ is not simple}
\STATE
 \textbf{Reduce} it to a simple graph 
 by identifying parallel edges.
\ENDIF

\STATE Sort the vertices into stacks, corresponding to
 degree 0, 1, 2, $\geq 5$, 4 and 3, in that order.

\STATE \textbf{Let} $G_0=G$.

\FOR{$i=1$ to $n$}
 \STATE \textbf{Pop} a next-reduction vertex 
  $v_i$ from the first non-empty stack. \label{pop}
 \IF{$\deg(v_i) \geq 5$}
  \STATE Check $v_i$ for duplicate incident edges. 
  \STATE Link any duplicate edge to the II-reduction that created it
  (using the label previously created by line~\ref{addmark}). 
 \ENDIF 
 \STATE \textbf{Reduce} $G_\im$ on $v_i$ to produce $G_i$,
  \emph{except}: \label{reductionline}
 \IF{$v_i$ had degree~2}
  \STATE \emph{Do not} check whether the added edge duplicates an existing one; 
  instead, \textbf{label} it as having been added 
  by the reduction on~$v_i$. \label{addmark}
 \ENDIF
 \STATE \textbf{Degree-check} each $G_\im$-neighbor of $v_i$.
   \label{A1degreecheck}
 \STATE \textbf{Place} each neighbor on the appropriate stack, 
  removing it from its old stack. 
\ENDFOR

\STATE \textbf{Output} the sequence $v_1,\dots,v_n$ of reduction vertices, 
 along with any duplicate-edge creations associated with 
 each II-reduction vertex.
 \label{A1output}
\end{algorithmic}
\end{algorithm*}

\begin{claim} \label{Aphase1}
On input of a graph $G$ with $n$ vertices and $m$ edges, 
\algaa 
runs in time and space $O(m+n)$ 
and produces a reduction sequence obeying the stated preference order.
\end{claim}

\begin{proof}
Correctness of the algorithm is guaranteed by line~\ref{pop}.
For the other steps we will have to detail some data structures 
and algorithmic details.

We assume a RAM model, so that a given memory location can be
accessed in constant time.
Let the input graph be presented in a sparse representation
consisting of a vector of vertices,
each with
a doubly-linked list of incident edges,
each edge with 
a pointer to the edge's twin copy indexed by the other endpoint.
 From the vector of vertices we create
a doubly linked list of them, so that
as vertices are removed from an instance to create a subinstance
they are bridged over in the linked list,
and there is always a linked list of just the vertices
in the subinstance.

Transforming the input graph into a simple one can be done in
time $O(m+n)$ and space $O(n)$. 
The procedure relies on a pointer array of length~$n$, initially empty.
For each vertex~$u$, we iterate through the incident edges. 
For an edge to vertex~$v$, if the $v$th entry of the pointer array is empty, 
we put a pointer to the edge~$uv$.
If the $v$th entry is not empty, this is not the first $uv$ edge we have seen,
and so we coalesce the new edge with the existing one:
using the pointer to the original edge, 
we use the link from the redundant $uv$ edge to its ``$vu$'' twin copy 
to delete the twin and bridge over it, 
then delete and bridge over the redundant $uv$ edge itself.
After processing the last edge for vertex~$u$ 
we run through its edges again, clearing the pointer array.
The time to process a vertex $u$ is of order the number of its incident edges
(or $O(1)$ if it is isolated),
so the total time is $O(m+n)$ as claimed.
Henceforth we assume without loss of generality that the 
input instance has no multiple edges. 

One of the trickier points is to 
maintain information about the degree of each vertex,
because a II-reduction may introduce multiple edges
and there is not time to run through its neighbors' edges 
to detect and remove parallel edges immediately. 
However, it will be possible to track whether each vertex 
has degree~0, 1, 2, 3, 4, or 5 or more.
We have a vertex ``stack'' for each of these cases. 
Each stack is maintained as a doubly linked list, 
and we keep pointers both ways between each vertex and
its ``marker'' in the stack.

The stacks can easily be \emph{created} in linear time from the input.
The key to \emph{maintaining} them
is a \emph{degree-checking} procedure for a vertex~$x$.
Iterate through $x$'s incident edges, 
keeping track of the number of distinct neighboring vertices seen, 
stopping when we run out of edges or find 5 distinct neighbors. 
If a neighbor is repeated, coalesce the two edges. 
The time spent on $x$ is $O(1)$ plus the number of edge coalescences.
Once the degree of $x$ is determined as 0, 1, 2, 3, 4, or 5 or more, 
remove $x$'s marker from its old stack 
(using the link from $x$ to delete the marker, 
and links from the marker to its predecessor and successor 
to bridge over it),
and push a marker to $x$ onto the appropriate new stack. 

When reducing on vertex~$v$, run the degree-checking procedure on 
each neighbor $x$ of~$v$
(line \ref{A1degreecheck} of \algaa).
The time for this is the time to count up to~5 for each neighbor
(a total of $O(\deg(v))$),
plus the number of edge coalescences. 
Vertex degrees never increase above their initial values, 
so over the course of \algaa the total of the $O(\deg(v))$ terms is~$O(m)$.
Parallel edges are created only by II-reductions, 
each producing at most one such edge, 
so over the course of \algaa at most $n$ parallel edges are created, 
and the edge coalescences thus take time~$O(n)$.
The total time for degree-checking is therefore~$O(m+n)$.

Finally, each reduction (line~\ref{reductionline} of \algaa)
can itself be performed in time~$O(1+\deg(v))$:
for a 0, I, or III-reduction we simply delete $v$ and its incident edges; 
for a II-reduction we do the same, then add an edge between $v$'s 
two former neighbors.
Again, the total time is $O(m+n)$.
\end{proof}

With \alga's first phase \algaa complete, 
we may assume henceforth that our graphs are always simple:
from this phase's output we can (trivially) reproduce the sequence 
of reductions in time $O(m+n)$, 
and coalesce any duplicate edge the moment it appears.

\subsection{Algorithm A: Second phase} \label{secAphase2}
The second phase, \algab, determines the optimum cost,
while the third and final phase, \algac, 
returns a coloring with this cost.
These two phases are nearly identical, and we proceed with \algab.

Because the algorithm is recursive and limited to linear space,
when recursing we cannot afford to pass a separate copy of the data; 
rather, a ``subinstance'' for recursion must be an in-place modification 
of the original data, 
and when a recursive call terminates it must restore 
the data to its original form.
This recursion is sketched in \algab (see displayed pseudocode).

\begin{algorithm}
\caption{\algab: \alga, second phase
recursively computing $s(G,S)$}
\label{A2A}
\begin{algorithmic}[1]
\STATE \textbf{Input:} 
 A CSP instance $(G,S)$, and
 reduction sequence $\vvec := v_1,\dots,v_n$
 (with associated duplicate-edge annotations, per \algaa
 line~\ref{A1output}).
\IF{$n=0$} 
 \STATE \textbf{Let} $s:=\snought$, the niladic score.
 \STATE \textbf{Return} $(s,G,S,\vvec)$.
\ENDIF 
\IF{$v_1$ is a 0-, I- or II-reduction vertex}
  \STATE \textbf{Reduce} $(G,S)$ on $v_1$ to obtain $(G',S')$
  \STATE \textbf{Record} an $O(r^2)$-space annotation allowing 
  the reduction on $v_1$ to be reversed. \label{a.2.9}
  \STATE \textbf{Truncate} the reduction sequence correspondingly, 
  letting $\vvec' := v_2,\dots,v_n$.
  \STATE \textbf{Let} $s:=s(G',S')$, computed recursively. \label{A2A-10}
  \STATE \textbf{Reverse} the reduction to reconstruct $(G,S)$ and $\vvec$
   (and free the storage from line~\ref{a.2.9}).
  \STATE \textbf{Return} $(s,G,S,\vvec)$.
\ELSE 
  \STATE $v$ is a III-reduction vertex.
  \STATE \textbf{Let} $s:=-\infty$.
  \FOR{color $C=1$ to $r$} \label{A2A-16}
   \STATE \textbf{III-reduce} on $v$ with color $C$ to obtain $(G',S^C)$, 
   and $\vvec'$. 
   \STATE \textbf{Record} an $O(\deg(v) r)$-space reversal annotation.  \label{a.2.18}
   \STATE \textbf{Let} $s:=\max\{s,s(G',S^C)\}$, computed recursively.
   \label{A2A-19}
   \STATE \textbf{Reverse} the reduction to reconstruct $(G,S)$
   (and free the storage from line~\ref{a.2.18}).
  \ENDFOR
  \STATE \textbf{Return} $(s,G,S,\vvec)$.
 \ENDIF
\STATE
 \textbf{Output:} $(s,G,S,\vvec)$, where $s$ is the optimal score of $(G,S)$.
\end{algorithmic}
\end{algorithm}

\begin{claim} \label{Aphase2}
Given an $(r,2)$-CSP instance 
with $n$ vertices, $m$ constraints, and length~$L$,
and a reduction sequence (per \algaa) with $\ell$ III-reductions,
\algab
returns the maximum score,
using space $O(L)$ and time $O(\oneplus r^{\ell+3} n)$.
\end{claim}

\begin{proof}
We first argue that each ``branch'' of the recursion
(determined by the colors chosen in the III-reductions)
requires space~$O(L)$. 

First we must detail how to implement the CSP reductions, 
which is a minor embellishment of the 
graph reduction implementations described earlier.
Recall that there is a score function on each vertex, 
which we will assume is represented as an $r$-value table, 
and a similar function on each edge, 
represented as a table with $r^2$ values.

A CSP II-reduction on $y$ with neighbors $x$ and $z$
follows the pattern of the graph reduction, 
but instead of simply constructing a new \emph{edge} $(x,z)$
we now construct a new \emph{score function}~$s'_{xz}$:
iterate through all color pairs $C,D \in [r]$
and set
$\ssp xzCD := \max_{F \in [r]} \{ \ss xyCF + \ss yzFD + \s yF\}$
as in~\eqref{IIreduction}. 
Iterating through values~$C$, $D$ and $F$ takes time~$O(r^3)$, 
and the resulting table takes space $O(r^2)$.
If there already was a score function $s_{xz}$
(if there already was an edge~$(x,z)$), 
the new score function is the elementwise sum of the two tables.
To reverse the reduction it suffices to record 
the neighbors $x$ and $z$ 
and keep around the old score functions 
$s_{xy}$ and $s_{yz}$ (allowing additional space~$O(r^2)$ for the new one).
Similarly, a I-reduction takes time $O(r^2)$ and space~$O(r)$, 
and a 0-reduction time $O(r)$ and space~$O(1)$.

To perform a III-reduction with color $C$ on vertex~$y$,
for each neighbor $x$ we incorporate the dyadic score 
$\ss yxCD$ into the monadic score~$\s xD$
(time $O(r)$ to iterate through $D \in [r]$),
maintain for purposes of reversal the original score functions 
$s_{yx}$ and $s_x$,
and allocate space $O(r)$ for the new score function $s'_x$.
Over all $\deg(y)$ neighbors the space required is $O(\deg(y) r)$,
and for each of the $r$ colors for the reduction, 
the time is also~$O(\deg(y) r)$.
(Note that $\deg(y)\neq 0$; indeed, $\deg(y)\geq 3$.)

Since vertex degrees are only decreased through the course of the algorithm, 
for one branch of the recursion the total space is 
$O(m r^2 + n r)$, \textit{i.e.},~$O(L)$.
Since each branch of the recursion takes space~$O(L)$, 
the same bound holds for the algorithm as a whole.

This concludes the analysis of space,
and we turn to the running time. 
Let $f(n,\ell)$ be an upper bound on the
running time for an instance
with $n$ nodes and III-recursion depth~$\ell$.
We claim that $f(0,0)=1$ and for $n>0$,
$f(n,\ell) \leq r^3 n (r^\ell+(r^{\ell+1}-r)/(r-1))$,
presuming that we have ``rescaled time'' so that all absolute constants
implicit in our $O(\cdot)$ expressions can be replaced by~1.
(This is equivalent to claiming that for some sufficiently large absolute
constant~$C$, $f(0,0)\leq C$ and
$f(n,\ell) \leq C r^3 n (r^\ell+(r^{\ell+1}-r)/(r-1))$.)
The case $n=1$ is trivial.
In the event of a recursive call in line~\eqref{A2A-10}, 
the recursion is preceded by just one 0-, I- or II-reduction, 
taking time $\leq r^3$;
the other non-recursive steps may also be accounted for in 
the same $r^3$ time bound.
By induction on~$n$, in this case we have
\begin{align*}
f(n,\ell) 
 & \leq r^3 + f(n-1,\ell)
 \\ & \leq r^3 + r^3 (n-1) (r^\ell+(r^{\ell+1}-r)/(r-1)) 
 \\ & \leq r^3 n (r^\ell+(r^{\ell+1}-r)/(r-1)) ,
\end{align*}
using only that 
$r^\ell+(r^{\ell+1}-r)/(r-1) \geq r^\ell \geq 1$.

The interesting case is where there are $r$ recursive 
calls originating in line~\eqref{A2A-19}, 
with the other lines in the loop~\eqref{A2A-16} 
consuming time $O(r \cdot \deg(v) r)$;
for convenience 
we bound this by $r^3 n$.
In this case, by induction on $n$ and~$\ell$,
\begin{align*}
f(n,\ell) & \leq 
r^3 n + r f(n-1,\ell-1)
 \\ &  \leq
 r^3 n + r \cdot r^3 n (r^{\ell-1}+(r^{\ell}-r)/(r-1))
 \\ & =
 r^3 n + r^3 n (r^{\ell}+(r^{\ell+1}-r^2)/(r-1))
 \\ & =
 r^3 n (r^{\ell}+(r^{\ell+1}-r^2 +r-1)/(r-1))
 \\ & \leq
 r^3 n (r^{\ell}+(r^{\ell+1}-r)/(r-1)) ,
\end{align*}
using $-r^2+r-1 \leq -r$ (from $0 \leq (r-1)^2$).
\end{proof}

\subsection{Algorithm A: Third phase} \label{secAphase3}

The third phase, \algac (not displayed)
proceeds identically to the second 
until we visit a leaf achieving the maximum score 
(known from the second phase),
at which point
we backtrack through all the reductions, filling in the vertex colors. 

There are two key points here.
The first is that when a maximum-score leaf is hit, we know it,
and can retrace back up the recursion tree.
The second is the property that in retracing up the tree, 
when we reach a node $v$,
all descendant nodes in the tree have been assigned optimal colors, 
$v$'s neighbors in the reduced graph correspond to such lower nodes,
and thus we can optimally color~$v$ (recursively preserving the property).
These points are obvious for \algac and so there is no need to 
write down its details,
but we mention them because \emph{neither property holds for \algb}, 
whose third phase \algbc is thus trickier.

Because \algac is basically just an interruption of \algab 
when a maximum-score leaf is encountered, 
the running time of \algac is no more than that of \algab. 
We have thus established the following claim.

\begin{claim} \label{Aresult}
Given an $(r,2)$-CSP instance 
with $n$ vertices, $m$ constraints, and length~$L$,
\alga returns an optimal score and coloring in 
space $O(L)$ and time $O(\oneplus r^{\ell+3} n)$, 
where $\ell$ is the number of III-reductions 
in the reduction sequence of \algaa.
\end{claim}

\subsection{Recursion depth} \label{Adepth}

The crux of the analysis is now to show that the number of III-reductions
$\ell$ in the reduction sequence produced by \alga's first phase 
is at most $m/5$.

\begin{lemma}
\label{depth5}
\algaa reduces a graph $G$ with $n$ vertices and $m$ edges to 
a vertexless graph
after no more than $m/5$ III-reductions.
\end{lemma}

\begin{proof}
While the graph has maximum degree 5 or more,
\alga III-reduces only on such a vertex,
destroying at least 5 edges;
any I- or II-reductions 
only increase
the number of edges destroyed.
Thus, it suffices to prove the lemma for graphs with maximum degree 4 or less.
Since the reductions never increase the degree of any vertex,
the maximum degree will always remain at most~4.

In this paragraph, we give some intuition for the rest of the argument.
\alga III-reduces on vertices of degree 4 as long as possible,
before III-reducing on vertices of degree~3, whose neighbors must then all be
of degree~3 (vertices of degree 0, 1 or 2 would trigger a 0-, I- or II-reduction
in preference to the III-reduction).
Referring to Figure~\ref{fig:reductions},
note that each III-reduction
on a vertex of degree~3 can be credited with destroying 6 edges,
if we immediately follow up with II-reductions on its neighbors.
(In \alga we do not explicitly couple the II-reductions to the III-reduction, 
but the fact that the III-reduction creates 3 degree-2 vertices 
is sufficient to ensure the good outcome that intuition suggests.  
In \algb we will have to make the coupling explicit.)
Similarly, reduction on a 4-vertex destroys at least 5 edges
unless the 4-vertex has no degree-3 neighbor.
The only problem comes from reductions on vertices of degree 4
all of whose neighbors are also of degree~4, 
as these destroy only 4 edges.
As we will see,
the fact that such reductions also create 4 3-vertices,
and the algorithm terminates with 0 3-vertices,
is sufficient to limit the number of times they are performed.
\newcommand{\fmath}[1]{{$#1$}}
\newcommand{\ind}{\hspace*{0.5cm}}
\begin{figure}
\begin{centering}
\hfill
\parbox[r]{2in}{%
\includegraphics[width=1.0in]{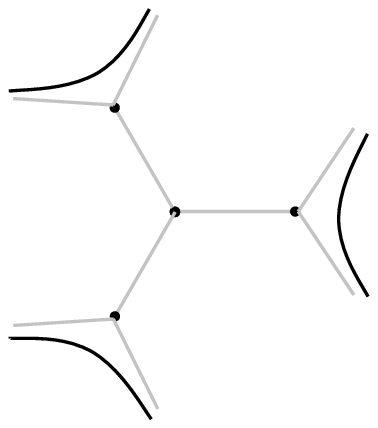}
}
\hfill
\parbox[l]{2in}{%
\includegraphics[width=1.0in]{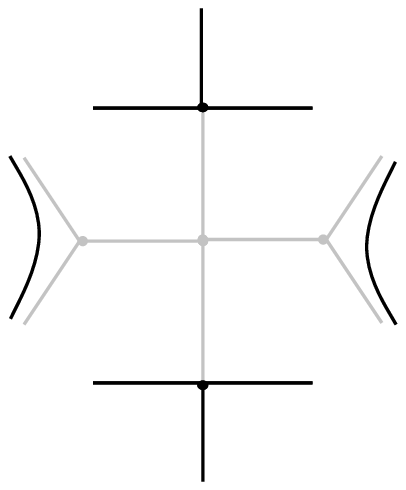}
}
\hfill
\end{centering}
\caption{
Left, 
a reduction on a 3-vertex with 3 3-neighbors,
followed by II-reductions on those neighbors,
destroys 6 edges and 4 3-vertices.
(The original graph's edges are shown in grey 
and the reduced graph's edges in black.)
Note that \alga does not actually force any particular II-reductions 
after a III-reduction, but \algb will do so.
Right, 
a reduction on a 4-vertex with $k$ 3-neighbors ($k=2$ here)
destroys $4+k$ edges and $2k-4$ 3-vertices
($k$ 3-vertices are destroyed, but $4-k$ 4-vertices become 3-vertices).
The algorithm and analysis make no assumptions
on the local structure of the graph; 
the figure is merely illustrative.
}
\label{fig:reductions}
\end{figure}

\long\def\tabrowfl #1 #2 #3 #4 #5%
 {$#1$ & #2 & #3 & #4 & #5 &}

\long\def\tabrowfr #1 #2 #3 #4 #5 #6%
 {$#1$ & $#2$ & $#3$ & $#4$ & $#5$ & #6 \\}

We proceed by considering the various types of reduction
and their effects on the number of edges and the number of 3-vertices.
The reductions are catalogued in Table~\ref{TableA}.

\begin{table}[htb]
\small
\begin{tabular}{|c|cccc||crrrr|c|}
\hline
\padhline
deg & \multicolumn{4}{c||}{\#nbrs of deg}
 & \multicolumn{5}{c|}{destroys} & steps \\
\tabrowfl \mbox{} 4 3 2 1 \tabrowfr e 4 3 2 1 {}
\hline \padhline
\tabrowfl 4 4 0 0 0 \tabrowfr 4 5 -4 0 0 1
\tabrowfl 4 3 1 0 0 \tabrowfr 4 4 -2 -1 0 1
\tabrowfl 4 2 2 0 0 \tabrowfr 4 3 0 -2 0 1
\tabrowfl 4 1 3 0 0 \tabrowfr 4 2 2 -3 0 1
\tabrowfl 4 0 4 0 0 \tabrowfr 4 1 4 -4 0 1
\tabrowfl 3 0 3 0 0 \tabrowfr 3 0 4 -3 0 1
\hline \padhline
\tabrowfl 2 {} {} {} {} \tabrowfr 1 0 0 1 0 0
\tabrowfl {\halfe} 1 0 0 0 \tabrowfr {\half} 1 -1 0 0 0
\tabrowfl {\halfe} 0 1 0 0 \tabrowfr {\half} 0 1 -1 0 0
\tabrowfl {\halfe} 0 0 1 0 \tabrowfr {\half} 0 0 1 -1 0
\tabrowfl {\halfe} 0 0 0 1 \tabrowfr {\half} 0 0 0 1 0
\hline
\end{tabular}
\padtableend
\caption{
Tabulation of the effects of various reductions in \alga.
}
\label{TableA}
\end{table}
The first row, for example, shows that III-reducing on a vertex of degree 4
with 4 neighbors of degree 4 (and thus no neighbors of degree~3)
destroys 4 edges, and (changing the neighbors from degree 4 to 3)
destroys 5 vertices of degree 4 (including itself)
and creates 4 vertices of degree~3.
It counts as one III-reduction ``step''.
The remaining rows up to the table's separating 
line similarly illustrate the other III-reductions.
Below the line, II-reductions and I-reductions are decomposed into parts.
As shown just below the line,
a II-reduction, regardless of the degrees of the neighbors,
first destroys 1 edge and 1 2-vertex,
and counts as 0 steps (steps count only III-reductions).
In the process, the II-reduction may create a parallel edge,
which will promptly be deleted (coalesced) by \alga.
Since the exact effect of an edge deletion 
depends on the degrees of its neighbors,
to minimize the number of cases
we treat an edge deletion as two half-edge deletions,
each of whose effects depends on the degree of the half-edge's incident vertex.
For example the table's next line 
shows deletion of a half-edge incident to a 4-vertex,
changing it to a 3-vertex and destroying half an edge.
The last four rows of the table also capture I-reductions.
0-reductions are irrelevant to the table, 
which does not consider vertices of degree~0.

The sequence of reductions reducing a graph to 
a vertexless graph
can be parametrized by an 11-vector $\nvec$ giving the number of
reductions (and partial reductions) indexed by the rows of the table,
so for example its first element is the number of
III-reductions on 4-vertices whose neighbors are also all 4-vertices.
Since the reductions destroy all $m$ edges,
the dot product of $\nvec$ with the table's column ``destroys e''
(call it $\evec$)
must be precisely~$m$.
Since all vertices of degree 4 are destroyed,
the dot product of $\nvec$ with the column ``destroys 4''
(call it $\df$) must be~$\geq 0$,
and the same goes for the ``destroy'' columns 3, 2 and~1.
The number of III-reductions is the dot product 
of $\nvec$ with the ``steps'' column,
$\nvec \cdot \steps$.
How large can the number of III-reductions $\nvec \cdot \steps$ possibly be?

To find out, let us maximize $\nvec \cdot \steps$
subject to the constraints that $\nvec \cdot \evec = m$
and that $\nvec \cdot \df$, $\nvec \cdot \dt$, $\nvec \cdot \db$ and $\nvec \cdot \dm$
are all $\geq 0$.
Instead of maximizing over proper reduction collections $\nvec$,
which seem hard to characterize,
we maximize over the larger class of non-negative real vectors $\nvec$,
thus obtaining an upper bound on the proper maximum.
Maximizing the linear function $\nvec \cdot \steps$ of $\nvec$
subject to a set of linear constraints 
(such as $\nvec \cdot \evec=m$ and $\nvec \cdot \df \geq 0$)
is simply solving a linear program (LP);
the LP's constraint matrix and objective function are the
parts of Table~\ref{TableA} right of the double line.
To avoid dealing with ``$m$'' in the LP,
we set $\nvec' = \nvec/m$, and
solve the LP with constraints $\nvec' \cdot \evec = 1$,
and as before $\nvec' \cdot \df \geq 0$, etc.,
to maximize $\nvec' \cdot \steps$.
The ``$\nvec'$'' LP is a small linear program 
(11 variables and 5 constraints)
and its maximum is precisely~$1/5$, 
showing that the number of III-reduction steps
--- $\nvec \cdot \steps = m \nvec' \cdot \steps$ ---
is at most $m/5$.

That the LP's maximum is at most 5 
can be verified from the LP's dual solution of 
$\yvec=(0.20, 0, -0.05, -0.2, -0.1)$.
It is easy to check that in each row, the ``steps'' value is
less than or equal to the dot product of this dual vector with the ``destroys'' values.
That is, writing $D$ for the whole ``destroys'' constraint matrix,
we have $\steps \leq D \yvec$.
Thus,
$\nvec' \cdot \steps
 \leq \nvec' \cdot (D \yvec)
 = (\nvec' D) \cdot \yvec$.
But $\nvec' D$ must satisfy the LP's constraints: 
its first element must be 1 and the remaining elements non-negative.
Meanwhile, the first element of $\yvec$ is $0.2$ 
and its remaining elements are non-positive,
so $\nvec' \cdot \steps \leq (\nvec' D) \cdot \yvec \leq 0.2$.
This establishes that the number of type-III reductions can be at most 
$1/5$th the number of edges~$m$, concluding the proof.
\end{proof}

\begin{theorem}
\label{thmA}
A \mc instance on $n$ variables with $m$ dyadic constraints 
and length $L$ can be solved
in time $O(\oneplus n r^{3 +m/5})$ and space~$O(L)$.
\end{theorem}

\begin{proof}
The theorem is an immediate consequence of 
Claim~\ref{Aresult} and Lemma~\ref{depth5}.
\end{proof}

The LP's \emph{dual} solution gives a ``potential function'' 
proof of Lemma~\ref{depth5}.
The dual assigns ``potentials'' 
to the graph's edges and to vertices according to their degrees, 
such that the number of steps counted for a reduction 
is at most its change to the potential.
Since the potential is initially at most $0.20 m$ and finally~0,
the number of steps is at most $m/5$.
(Another illustration of duality appears in the proof
of Lemma~\ref{lpalpha}.)

The \emph{primal} solution of the LP, which describes the worst case,
uses (proportionally) 1 III-reduction on a 4-vertex with all 4-neighbors,
1 III-reduction on a 3-vertex,
and 3 II-reductions
(the actual values are 1/10th of these).
As it happens, this LP worst-case bound is achieved by the 
complete graph~$K_5$,
whose 10 edges are destroyed by two III-reductions and then
some I- and II-reductions.

\section{An \timeb algorithm}
\label{secb}

\newcommand{\badv}{bad\xspace}
\newcommand{\badq}{$(4|040)$\xspace}
\newcommand{\goodq}{$(4|004)$\xspace}
\newcommand{\badp}{$(5|500)$\xspace}
\newcommand{\goodp}{$(5|005)$\xspace}
\renewcommand{\d}{\nvec}

\subsection{Improving \alga}
The analysis of \alga contains the seeds of its improvement. 
First, since reduction on a 5-vertex may destroy only 5 edges, 
we can no longer ignore such reductions if we want to improve on~$m/5$. 
This simply means including them in the LP.

Second, were this the only change we made, we would find the 
LP solution to be the same as before
(adding new rows leaves the previous primal solution feasible).
The solution
is supported on a ``bad'' reduction destroying only 4 edges
(reducing on a 4-vertex with all 4-neighbors), while the other
reductions it uses are more efficient.
This suggests that we should focus on eliminating the bad reduction. 
Indeed, if in the LP we ascribe 0 ``steps'' to the bad reduction instead of~1, 
the LP cost decreases to $23/120$ (about $0.192$),
and support of the new solution includes reductions on a 
degree-5 vertex with all degree-5 neighbors
and on a degree-4 vertex with one degree-3 neighbor,
each resulting in the destruction of only 5 edges.
Counting 0 steps instead of 1 for this degree-5 reduction 
gives the LP a cost of~$19/100$, 
suggesting that if we could somehow avoid this reduction too, 
we might be able to hope for an algorithm running in time
$\Ostar{r^{19m/100}}$; 
in fact our algorithm will achieve this.
Further improvements could come from avoiding the next bad cases
---
a 5-vertex with neighbors of degree 5 except for one of degree~4,
and a 4-vertex with neighbors of degree 4 except for one of degree~3
---
but we have not pursued this.

Finally, we will also need to take advantage of the component structure of our graph.
For example, a collection of many disjoint $K_5$ graphs 
requires $m/5$ III-reductions in total. 
To beat $\Ostar{r^{m/5}}$ we will have to use the fact that
an optimum solution to a disconnected CSP is a union of solutions
of its components, 
and thus that the $m/5$ reductions can in some sense be done in parallel, 
rather than sequentially.
Correspondingly, where \alga built a \emph{sequence} of reductions 
of length at most $m/5$,
\algb will build a \emph{reduction tree} whose III-reduction \emph{depth} 
is at most $2+19m/100$.
The depth bound is proved by showing 
that in any sequence of reductions in a component on a fixed vertex, 
all but at most two ``bad'' reductions can be paired with other reductions, 
and for the good reductions (including the paired ones), 
the LP has maximum~$19/100$.

\subsection{\algb: General description}

Like \alga, 
\algb preferentially performs type 0, I or II reductions,
but it is more particular about the vertices on which it III-reduces.
When forced to perform a type III reduction, \algb
selects a vertex in the following decreasing order of preference:
\begin{itemize}
\item
a vertex of degree~$\geq 6$;
\item
a vertex of degree~5 with at least one neighbor of degree 3 or~4;
\item
a vertex of degree~5 whose neighbors all have degree~5;
\item
a vertex of degree 4 with at least one neighbor of degree~3;
\item
a vertex of degree 4 whose neighbors all have degree~4;
\item
a vertex of degree~3.
\end{itemize}
When \algb makes any such reduction with any degree-3 neighbor,
it immediately follows up with II-reductions on all those neighbors.%
\footnote{An example of this was shown in Figure~\ref{fig:reductions}.  In some
cases, we may have to use I-reductions or 0-reductions 
instead of II-reductions (for instance if the degree-3
neighbors contain a cycle), but
the effect is still to destroy one edge and one vertex 
for each degree-3 neighbor.%
}  
\algb then recurses separately on each component of the resulting graph.

As before, in order to get an efficient implementation 
we must be careful about details.  
Section~\ref{bfirst} discusses the construction of the ``reduction tree''; 
a reader only interested in 
an $\Ostar{r^{19m/100}}$ bound could skip Lemma~\ref{treeBlemma} there.  
Section~\ref{breduct} is essential, 
and gives the crucial bound 
$19m/100 + O(1)$ on the depth of a reduction tree, 
while Section~\ref{secB2} establishes that if the depth of a reduction tree 
is $d$ then an optimal score can be found in time $\Ostar{r^d}$.  
Finally, Section~\ref{sec3B} ties up loose ends 
(including how to move from an optimal score to an optimal assignment) 
and gives the main result of this part of the paper 
(Theorem~\ref{maintheorem}).

\subsection{\algb: First phase}\label{bfirst}

As with \alga, a first phase \algba of \algb 
starts by identifying a sequence of graph reductions.
Because \algb will treat graph components individually, 
\algba then organizes this \emph{sequence} of reductions into a 
\emph{reduction tree}.
The tree has vertices in correspondence with those of~$G$,
and the defining property 
that if reduction on a (graph) vertex $v$ divides the 
graph into $k$ components, 
then the corresponding tree vertex $v$ has $k$ children, 
one for each component, where
each child node corresponds to the first vertex reduced upon 
in its component 
(\textit{i.e.}~the first vertex in the reduction sequence restricted to 
the set of vertices in the component).
If the graph is initially disconnected, the reduction ``tree'' 
is really a forest, but since this case presents no additional issues 
we will speak in terms of a tree.
We remark that the number of child components $k$ 
is necessarily 1 for I- and II-reductions, 
can be 1 or more for a III-reduction, 
and is 0 for a 0-reduction.

We define the \emph{III-reduction depth} of an instance to be the
maximum number of III-reduction nodes in any 
root-to-leaf path in the reduction tree.
Lemma~\ref{treeBlemma} characterizes an efficient construction of the tree, 
but it is clear that it can be done in polynomial time and space. 
The crux of the matter is Lemma~\ref{depth19}, 
which relies on the reduction preference order set forth above,
but not on the algorithmic details of \algba.

\begin{algorithm}
\caption{\algba: \algb, first phase}
\begin{algorithmic}[1]
\STATE
 \textbf{Input} a constraint graph $G_0=G$.
\FOR{
 $i=1$ {to} $n$ }
\STATE
 \textbf{Select} a vertex $v_i$ by the preference order described above.
\STATE
 \textbf{Reduce} 
 $G_{i-1}$ on $v_i$ to produce $G_i$.
\ENDFOR
\STATE
 \textbf{Initialize} $T$ to be an empty forest.
\FOR{
 $i=n$ {to} $1$ }
\STATE
 \textbf{Reverse} the $i$th reduction. 
 For a 0-reduction on $v_i$, 
 add an isolated node $v_i$ to the forest~$T$.
 For a I-reduction on $v_i$ with neighbor $x_i$ in $G_\im$,
 set $v_i$ to be the parent of the root node of the component of $T$
 containing~$x_i$.
 Do the same for a II-reduction on $v_i$, whose $G_\im$-neighbors 
 $x_i$ and $y_i$ will belong to a common component of~$T$.
 For a III-reduction on $v_i$, unite all component trees of $T$ 
 containing $G_\im$-neighbors of $v_i$ by setting $v_i$ as the common 
 parent of their roots.
 (Details of an efficient implementation are in the proof of Lemma~\ref{treeBlemma}.)
\ENDFOR
\STATE
 \textbf{Output}
 the reduction tree~$T$. 
 $T$ has the property that for any node $v \in T$, 
 reducing $G$ on all ancestor nodes of the corresponding node $v \in G$
 produces a graph $G'$ whose component containing $v$ has vertex set equal 
 to the vertex set of the subtree of $T$ rooted at~$v$.
\end{algorithmic}
\end{algorithm}

\begin{lemma} \label{treeBlemma}
A reduction tree on $n$ vertices which has III-reduction depth $d$
can be constructed in time $O(d n+n)$ and space~$O(m+n)$.
\end{lemma}

\begin{proof}
We use \algba (see displayed pseudocode).
First the sequence of reductions is found much as in \algaa
and in the same time and space
(see Claim~\ref{Aphase1}).
As long as there are any vertices of degree $\geq 6$ this 
works exactly as in \algaa, but with stacks up to degree~6.
Once the degree-6 stack is empty it will remain empty 
(no reduction increases any vertex degree) 
and at this point we create stacks according to the 
degree of a vertex and the degrees of its neighbors
(for example, a stack for vertices of degree 5 with two neighbors of 
degree 5 and one neighbor each with degrees 4, 3 and 2). 
Since the degrees are bounded by 5 this is a small constant number of stacks, 
which can be initialized in linear time.
After that, for each vertex whose degree is affected by a reduction
(and which thus required processing time $\Omega(1)$ in \algaa), 
we must update the stacks for its at most 5 neighbors
(time $O(1)$); 
this does not change the complexity.

To form the reduction tree we read 
backwards through the sequence of reductions
growing a collection of subtrees, starting from the leaves, 
gluing trees together into larger ones when appropriate,
and ending with the final reduction tree.
We now describe this in detail, and analyze the time and space of \algba.

Remember that there is a direct correspondence between reductions, 
vertices of the CSP's constraint graph, 
and nodes in the reduction tree.
At each stage of the algorithm 
we have a set of subtrees of the reduction tree, 
each subtree labeled by some vertex it contains.
We also maintain a list which indicates, for each vertex, 
the label of the subtree to which it belongs,
or ``none'' if the corresponding reduction has not been reached yet.
Finally, for each label, there is a pointer to the corresponding tree's root.

Reading backwards through the sequence of reductions,
we consider each type of reduction in turn.
\begin{description}
\item[0-reduction]
A forward 0-reduction on $y$ destroys the isolated vertex~$y$, 
so the reverse reduction creates a component consisting only of~$y$.
We create a new subtree consisting only of~$y$,
label it ``$y$'', root it at~$y$,
and record that $y$ belongs to that subtree.
\item[I-reduction]
If we come to a I-reduction on vertex $y$ with neighbor~$x$,
note that $x$ must already have been seen in our backwards reading 
and, since I-reductions do not divide components,
the reversed I-reduction does not unite components. 
In this case we identify the tree to which $x$ belongs,
leave its label unchanged,
make $y$ its new root, 
make the previous root $v$ (typically $v \neq x$) the sole child of~$y$,
update the label root-pointer from $v$ to~$y$,
and record that $y$ belongs to this tree.
\item[II-reduction]
For a II-reduction on vertex $y$ with neighbors $x$ and~$z$,
the forward reduction merely replaces the $x$--$y$--$z$ path with 
the edge $x$--$y$ and thus does not divide components.
Thus the reversed reduction does not unite components,
and so in the backwards reading 
$x$ and $z$ must already belong to a common tree.
We identify that tree,
leave its label unchanged,
make $y$ its new root, 
make the previous root the sole child of~$y$,
and record that $y$ belongs to this tree.
\item[III-reduction]
Finally, given a III-reduction on vertex~$y$, 
we consider $y$'s neighbors $x_i$,
which must previously have been considered in the backwards reading.
We unite the subtrees for the $x_i$ into a single tree with root~$y$,
$y$'s children consisting of 
the roots for the labels of the~$x_i$.
(If some or all the $x_i$ already belong to a common subtree,
we take the corresponding root just once.
Since the roots are values between $1$ and~$n$, 
getting each root just once can be done without any increase in complexity
using a length-$n$ array; 
this is done just as we eliminated parallel edges on a vertex 
in \alga's first phase --- see the proof of Claim~\ref{Aphase1}.)
We give the resulting tree the new label~$y$,
abandon the old labels of the united trees, 
and point the label $y$ to the root~$y$.
Relabeling the tree also means
conducting a depth-first search to find all the tree's nodes
and update the label information for each.
If the resulting tree has size $n'$ 
the entire process takes time~$O(n')$.
\end{description}

In the complete reduction tree,
define ``levels'' from the root based only on nodes corresponding
to III-reductions
(as if contracting out nodes from 0, I and II-reductions).
The III-reduction nodes at a given level of the tree have disjoint subtrees,
and thus 
in the ``backwards reading''
the total time to process all of these nodes together is~$O(n)$.
Over $d$ levels, this adds up to~$O(d n)$.
The final time bound $O(dn+n)$ 
also accommodates time to process all $O(n)$ 0-, I- and II-reductions.

The space requirements are a minimal~$O(n)$: 
beyond the space implicit in the input and that entailed 
by the analog of \algaa, 
the only space needed is the $O(n)$ to maintain the labeled forest.
\end{proof}

\subsection{Reduction-tree depth}\label{breduct}

Analogous to Lemma~\ref{depth5} characterizing \alga,
the next lemma is the heart of the analysis of \algb.

\begin{lemma} \label{depth19}
For a graph $G$ with $m$ edges,
the reduction tree's III-reduction depth is $d \leq 2+ 19m/100$.
\end{lemma}

\begin{proof}
By the same reasoning as in the proof of Lemma~\ref{depth5},
it suffices to prove the lemma for graphs with maximum degree at most~5.

Define a ``bad'' reduction to be one
on a 5-vertex all of whose neighbors are also of degree~5,
or on a 4-vertex all of whose neighbors are of degree~4.
(These two reductions destroy 5 and 4 edges respectively, while
most other reductions, coupled with the II-reductions they enable,
destroy at least 6 edges.)
The analysis is aimed at controlling the number of bad reductions.
In particular, we show that every occurrence of a bad
reduction can be paired with one or more ``good'' reductions, 
which delete enough edges to compensate for the bad reduction.

For shorthand, we write reductions in terms of the degree of the vertex
on which we are reducing followed by the numbers of neighbors of
degrees 5, 4, and~3,
so for example the bad reduction on a 5-vertex is written~\badp.
Within a component, a \badp reduction is performed only if there
is no 5-vertex adjacent to a 3- or 4-vertex;
this means the component \emph{has} no 3- or 4-vertices,
since otherwise a path from such a vertex to the 5-vertex
would include an edge incident on a 5-vertex and a 3- or 4-vertex.

We bound the depth by tracking
the component containing a fixed vertex, say vertex~1,
as it is reduced.
Of course the same argument 
(and therefore the same depth bound) applies to every vertex.
If the component necessitates a ``bad 5-reduction''
(a bad III-reduction on a vertex of degree~5),
one of four things must be true:
\label{casetable}
\begin{description}
\item[C1]
This is the first degree-5 reduction in this branch of the reduction tree.
\item[C2]
The previous III-reduction 
(the first III-reduction ancestor in the reduction tree,
which because of our preference order 
must also have been a degree-5 reduction) 
was on a $(5|005)$ vertex, and left no vertices of degree 3 or~4.
\item[C3]
The previous III-reduction was on a 5-vertex and produced
vertices of degree 3 or 4 in this component,
but they were destroyed by I- and II-reductions.
\item[C4]
The previous III-reduction was on a 5-vertex and produced
vertices of degree 3 or~4,
but split them all off into other components.
\label{splitcase}
\end{description}

As in the proof of Lemma~\ref{depth5}, for each type of reduction
we will count:
its contribution to the depth
(normally~1 or 0, but we also introduce ``paired'' reductions 
counting for depth~2);
the number of edges it destroys;
and
the number of vertices of degree 4, 3, 2, and 1 it destroys.
Table~\ref{TableB} shows this tabulation.
In \algb we immediately follow each III-reduction with
a II-reduction on each 2-vertex it produces,
so for example in row~1 a $(5|005)$ reduction destroys a total of 10 edges
and 5 3-vertices; 
it also momentarily creates 5 2-vertices but immediately reduces them away.

The table's boldfaced rows and the new column ``forces'' require explanation.
They relate to the elimination of the bad \badp reduction from the table,
and its replacement 
with versions corresponding to the cases above.

Case (C1) above can occur only once.
Weakening this constraint, we will allow it to occur any number
of times, but we will count its depth contribution as~0,
and add 1 to the depth at the end.
For this reason, the first bold row in Table~\ref{TableB} has depth 0 not~1.

In case (C2) we may pair the bad \badp reduction with
its preceding \goodp reduction.
This defines a new ``pair'' reduction
shown as the second bold row of the table:
it counts for 2 steps,
destroys 15 edges, etc.
(Other, non-paired \goodp good reductions are still allowed as before.)

In case (C3) we wish to similarly pair the \badp reduction with
a I- or II-reduction, but we cannot say specifically with
which sort. 
The ``forces'' column of Table~\ref{TableB}
will constrain each \badp reduction for this case to be accompanied by at least
one I-reduction (two half-edge reductions of any sort) or II-reduction.

In case~(C4), the \badp reduction produces a non-empty side component
destroyed with the usual reductions but adding depth~0 
to the component of interest.
These reductions can be expressed as
a nonnegative combination of half-edge reductions, 
which must destroy at least one edge, 
so we force the \badp reduction to be accompanied by 
at least two half-edge reductions, precisely as in case~(C3).
Thus case~(C4) does not require any further changes to the table.

Together, the four cases mean that we were able to 
exclude \badp reductions,
replacing them with less harmful possibilities represented
by the first three bold rows in the table.

We may reason identically for bad $(4|040)$ reductions on 4-vertices,
contributing the other three bold rows. 
We reiterate the observation that I-reductions,
as well as the merging of parallel edges,
can be written as a nonnegative combination of half-edge reductions.

\begin{table}[htb] 
\small
\begin{tabular}{|r@{\hspace*{0.4cm}}||c|ccccc||crrrr|r|c|}
\hline \padhline
line \# & deg & \multicolumn{5}{c||}{\#nbrs of deg}
 & \multicolumn{5}{c|}{destroys} & forces & depth \\
\tabrownonum  \mbox{} 5 4 3 2 1  {$e$} 4 3 2 1 { } { } 
\hline \hline \padhline
\tabrow    5  0  0  5  0  0 10  0  5  0  0   0   1
\tabrow    5  0  1  4  0  0  9  1  3  0  0   0   1
\tabrow    5  0  2  3  0  0  8  2  1  0  0   0   1
\tabrow    5  0  3  2  0  0  7  3 -1  0  0   0   1
\tabrow    5  0  4  1  0  0  6  4 -3  0  0   0   1
\tabrow    5  0  5  0  0  0  5  5 -5  0  0   0   1
\tabrow    5  1  0  4  0  0  9 -1  4  0  0   0   1
\tabrow    5  1  1  3  0  0  8  0  2  0  0   0   1
\tabrow    5  1  2  2  0  0  7  1  0  0  0   0   1
\tabrow    5  1  3  1  0  0  6  2 -2  0  0   0   1
\tabrow    5  1  4  0  0  0  5  3 -4  0  0   0   1
\tabrow    5  2  0  3  0  0  8 -2  3  0  0   0   1
\tabrow    5  2  1  2  0  0  7 -1  1  0  0   0   1
\tabrow    5  2  2  1  0  0  6  0 -1  0  0   0   1
\tabrow    5  2  3  0  0  0  5  1 -3  0  0   0   1
\tabrow    5  3  0  2  0  0  7 -3  2  0  0   0   1
\tabrow    5  3  1  1  0  0  6 -2  0  0  0   0   1
\tabrow    5  3  2  0  0  0  5 -1 -2  0  0   0   1
\tabrow    5  4  0  1  0  0  6 -4  1  0  0   0   1
\tabrow    5  4  1  0  0  0  5 -3 -1  0  0   0   1
\boldtabrow    5  5  0  0  0  0  5 -5  0  0  0   0   0
\boldtabrow   5+5  5  0  5  0  0 15 -5  5  0  0   0   2
\boldtabrow    5  5  0  0  0  0  5 -5  0  0  0  -1   1
\tabrow    4  0  0  4  0  0  8  1  4  0  0   0   1
\tabrow    4  0  1  3  0  0  7  2  2  0  0   0   1
\tabrow    4  0  2  2  0  0  6  3  0  0  0   0   1
\tabrow    4  0  3  1  0  0  5  4 -2  0  0   0   1
\boldtabrow    4  0  4  0  0  0  4  5 -4  0  0   0   0
\boldtabrow    4+4  0  4  4  0  0 12  6  0  0  0   0   2
\boldtabrow    4  0  4  0  0  0  4  5 -4  0  0  -1   1
\tabrow    3  0  0  3  0  0  6  0  4  0  0   0   1
\hline \padhline
\newcounter{line3}
\setcounter{line3}{\value{rownum}}
\tabrow    2  0  0  0  0  0  1  0  0  1  0   1   0
\tabrow    {\halfe}  1  0  0  0  0  {\half} -1  0  0  0   {\half}   0
\tabrow    {\halfe}  0  1  0  0  0  {\half}  1 -1  0  0   {\half}   0
\tabrow    {\halfe}  0  0  1  0  0  {\half}  0  1 -1  0   {\half}   0
\tabrow    {\halfe}  0  0  0  1  0  {\half}  0  0  1 -1   {\half}   0
\tabrow    {\halfe}  0  0  0  0  1  {\half}  0  0  0  1   {\half}   0
\hline
\end{tabular}
\padtableend
\caption{
Tabulation of the effects of various reductions in \algb.
}
\label{TableB}
\end{table}

In analyzing a leaf of the reduction tree, 
let vector $\nvec$ count the number of reductions of each type,
as in the proof of Lemma~\ref{depth5}.
As before, the dot product of $\nvec$ with the ``destroys $e$'' column
is constrained to be~1
(we will skip the version where it is $m$ and go straight to the normalized form),
its dot products with the other ``destroys'' columns must be non-negative,
ditto its dot product with the ``forces'' column,
and the question is how large
its dot product $x$ with the ``depth'' column can possibly be.
For then,
unnormalizing, the splitting-tree depth of vertex 1 as we counted it is
at most $x m$, and the true III-reduction depth
(accounting for the possible case~(C1) occurrences for 4- and 5-vertices)
is at most $2+x m$.

As before, $x$
is found by solving the LP: it is~$19/100$.
The dual solution, with weights
$(0.190$, $-0.005$, $-0.035$, $0$, $0$, $0.150)$
on edges, degrees 4, 3, 2, 1, and ``forces'',
witnesses this as the maximum possible.
(For more on duality, see the proof of Lemma~\ref{lpalpha}.)
This concludes the proof.
\end{proof}

We observe that the maximum is achieved by a weight vector with just
three nonzero elements, putting relative weights of 8, 6, and 5 on
the reductions $(5|410)$, $(4|031)$, and $(3|003)$. 
That is, the proof worked by essentially eliminating bad reductions
of types \badp and \badq
(which destroy only 5 and 4 edges respectively,
in conjunction with the II-reductions they enable),
and the bound produced uses the second-worst reductions,
of types $(5|410)$ and $(4|031)$
(each destroying 5 edges, with the accompanying II-reductions),
which it is forced to balance out with 
favorable III-reductions of type $(3|003)$.

\begin{remark} \label{depth4}
For an $m$-edge graph $G$ and maximum degree $\leq 4$,
the reduction tree's III-reduction depth is $d \leq 1+ (3/16) m$.
If $G$ has maximum degree $\leq 3$, the depth is $d \leq m/6$.
\end{remark}

\begin{proof}
The first statement's proof is identical to that of Lemma~\ref{depth19} 
except that from Table~\ref{TableB} we discard reductions (rows)
involving vertices of degree~5, 
we solve the new LP,
and we have an additive 1 instead of~2
(for a single bad reduction on a vertex of degree~4, 
rather than one each for degrees 4 and~5).
The second statement can be obtained directly and trivially,
or we may
go through the same process.
\end{proof}

\subsection{\algb: Second phase} \label{secB2}

It is straightforward to compute the optimal \emph{score} of an instance;
this is \algbb (see displayed pseudocode).
\begin{algorithm}
\caption{\algbb: \algb, second phase}
\begin{algorithmic}[1]
\STATE
\textbf{Input:} 
 The input consists of a CSP instance $(G,S)$, 
 a tree~$T$, and a vertex $v \in T$ such that the 
 subtree of $T$ rooted at $v$ is a reduction tree for the 
 component of $(G,S)$ containing~$v$.
 (We start with an initial CSP $(G_0,S_0)$ 
 with reduction tree~$T$,
 and $(G,S)$ is the reduction of $(G_0,S_0)$ on the ancestors
 of~$v$, with some choices of colors for the III-reductions.)
\STATE \textbf{Let} $v'$ be the first 0- or III-reduction node 
below (or equal to)~$v$.
\STATE \textbf{I- and II-reduce} on all nodes from $v$ 
up to but not including~$v'$.
(If $v=v'$, do nothing.)
\IF{$v'$ is a 0-reduction node}
  \STATE Reduce on $v'$ and return the resulting niladic score~$s$.
\ENDIF
\STATE Let $v_1,\dots,v_k$ be the children of $v'$. Let $s:=-\infty$.
\FOR{color $C=1$ to $r$}
  \STATE \textbf{III-reduce} on $v'$ with color~$C$.
  \STATE Let $s':=0$.
  \FOR{$i=1,\dots,k$}
    \STATE Let $s':= s'+ \bb(v_i)$, computed recursively.
  \ENDFOR
  \STATE Let $s:= \max\{s,s'\}$, 
\ENDFOR
\STATE
 \textbf{Output:} $s$, the optimal score of the component of $G$ 
 containing~$v$.
\end{algorithmic}
\end{algorithm}
As with \algab, \algbb is a recursive procedure which,
with the exception of a minimal amount of state information, 
works ``in place'' in the global data structure 
for the problem instance. 
In addition to the algorithm's explicit input, 
state information is a single active node $\vstar \in T$
(a descendant of~$v$), and, for each ancestor of~\vstar: 
a reference to which of its children leads to \vstar; 
the sum of the optimal scores for the earlier children; 
its current color;
and the usual information needed to reverse the reduction.

The recursion can be executed with a global state consisting of a 
path from the root node to the currently active node, 
along with a color for each III-reduction node along the path:
after the current node $\vstar$ and color have been explored, 
if possible the color is incremented, 
otherwise if there is a next sibling of $\vstar$ it is tried with color~1, 
otherwise control passes to the first III-reduction ancestor of $v'$, 
and if there is no such ancestor then the recursion is complete.

Define the depth $d$ of a tree node $v$ to be the maximum, 
over all leaves $\ell$ under~$v$, 
of the number of III-reduction nodes from $v$ to $\ell$ inclusive.
The following claim governs the running time of~\algbb.

\begin{claim} \label{bb}
For a tree node $v$ of depth $d$ whose subtree has order $n_v$,
\algbb runs in time $O(n_v r^{3+d})$
and in linear space.   
\end{claim}

\begin{proof}
Any sequence of 0-, I- and II-reductions 
can be performed in time $O(r^3 n)$, 
and a set of $r$ III-reductions (one for each color) 
in time $O(r^2 n)$
(see the proof of Claim~\ref{Aphase2}).
Let us ``renormalize'' time so that the sum of these two can be 
bounded simply by $r^3 n$ 
(again as in the proof of Claim~\ref{Aphase2}).
We will prove by induction on $d$ that 
an instance of order $n_v$ and depth $d$ can be solved in time at most
\begin{align}
  f(n_v,d) & := r^3 n_v \big(r^d + (r^{d+1}-r)/(r-1) \big) ,
\label{recurbound}
\end{align}
which is at most $3 n_v r^{3+d}$.

The base case is that $d=0$, no III-reductions are required,
and the instance is solved by performing and reversing
a series of 0-, I- and II-reductions;
this takes time $\leq r^3 n_v$,
which is smaller than the right-hand side of~\eqref{recurbound}. 

For a node $v$ of depth $d>0$,
define $v'$ to be the first III-reduction descendant of $v$ 
(or $v$ itself if $v$ is a III-reduction node).
The reductions from $v$ up to but not including $v'$, 
and the $r$ possible reductions on $v'$,
take time $\leq r^3 n$. 
The total time taken by \algbb is this plus the time to 
recursively solve each of the $r$ subinstances reduced from~$v'$. 
If the tree node $v'$ has outdegree~$k$, 
each of the $r$ subinstances decomposes 
into $k$ components, the $i$th component having order $n_i$ and depth $d_i$
(with $n_1+\dots+n_k=n_v-1$, and $d_i \leq d-1$),
and thus the total time taken is 
$f(n_v,d) \leq r^3 n + r \sum_{i=1}^k f(n_i,d_i)$.
By the inductive hypothesis~\eqref{recurbound}, then,
{
\allowdisplaybreaks
\begin{align*}
f(n_v,d) & \leq 
r^3 n_v + r \sum_{i=1}^k f(n_i,d_i)
 \\ &  \leq
 r^3 n_v + r \sum_{i=1}^k r^3 n_i \big(r^{d_i} + (r^{d_i+1}-r)/(r-1) \big)
 \\ & \leq
 r^3 n_v + r \sum_{i=1}^k r^3 n_i \big(r^{d-1} + (r^{d}-r)/(r-1) \big)
 \\ & <
 r^3 n_v + (r^3 n_v) r \big(r^{d-1} + (r^{d}-r)/(r-1) \big)
 \\ & \leq
 r^3 n_v \big(1+ r^{d} + (r^{d+1}-r^2)/(r-1) \big)
 \\ & \leq
 r^3 n_v \big(r^{d} + (r^{d+1}-r)/(r-1) \big) .
\end{align*}
}

The linear space demand follows just as for \algab.
\end{proof}

\subsection{\algb: Third phase}\label{sec3B}
In \alga, the moment an optimal score is achieved
(at the point of reduction to an empty instance), 
all III-reduction vertices already have their optimal colors,
and reversing all reductions gives an optimal coloring.
This approach does not work for \algb, 
because we now have a tree of reductions 
rather than a path of reductions.

Imagine, for example, 3-coloring a III-reduction vertex $A$ with 
children $B$ and $C$ that are also III-reduction vertices, 
and where the optimal colors are $\phi(A)=1$, $\phi(B)=2$, $\phi(C)=3$.
We first try the coloring $\phi(A)=1$, and within this  
we try the six (not~nine!) combinations $\phi(B)=1,2,3$
and then $\phi(C)=1,2,3$.
Even knowing the optimal score, there is no ``moment of truth'' 
when the score
is achieved: we have gone past $\phi(B)=2$ by the 
time we start with $\phi(C)=1$.
Also, even if we could recover the fact that for $\phi(A)=1$ the optimal
settings were $\phi(B)=2$, $\phi(C)=3$, 
we would not be able to remember this as we were trying $\phi(A)=2,3$. 
(In this simple example we would already be forced to remember optimal 
choices for both $B$ and $C$ for each possible color of~$A$,
and taking the full tree into account this would 
become an exponential memory requirement.)

Fortunately, there is a relatively simple work-around. 
Having computed the optimal score with \algbb, 
we can try different colors at the highest III-reduction vertex 
to see which gives that score; 
this gives the optimal coloring of that vertex.
(It is worth noting that we cannot immediately reverse the ancestor
I- and II-reductions, as those vertices may be adjacent to vertices 
not yet colored; coloring by reversing reductions only works
after we have reduced to an empty instance.)
We can repeat this procedure, working top down, 
to optimally color all III-reduction vertices.
After this, it is trivial to color all the remaining, 
0-, I- and II-reduction vertices.
These stages are all described as \algbc
(see displayed pseudocode).

\begin{algorithm}[tb]
\caption{\algbc: \algb, third phase}
\begin{algorithmic}[1]
\STATE
 \textbf{Input} a CSP $(G,S)$ and a reduction tree $T$ for~$G$.
\FOR{each III-reduction node $v \in T$, in 
depth-first search
order (by first visit)}
    \STATE \textbf{Let} $s:=\bb(v)$. \label{b.c.3}
    \STATE \textbf{Let} $v_1,\ldots,v_k$ be the children of~$v$. 
    \FOR{color $C=1$ to $r$}
      \STATE \textbf{III-reduce} on $v$ with color~$C$.
      \IF{$s= \bb(v_1)+\dots+\bb(v_k)$} \label{b.c.7}
        \STATE Assign color $C$ to $v$ and \textbf{break}.
      \ENDIF
    \ENDFOR
\ENDFOR
\STATE 
At this point all III-reduction nodes of $G$ are colored, optimally.
\STATE
\textbf{Perform} all corresponding III-reductions on $G$, 
using these optimal colors,
to derive an equivalent instance~$G'$. 
\STATE
\textbf{Perform} the 0-, I- and II-reductions of~$T$, in 
depth-first search order,
reducing $G'$ to an empty instance.
\STATE
\textbf{Reverse} the 0-, I- and II-reductions 
to optimally color all vertices of $G'$, and thus of~$G$.
\STATE
 \textbf{Output} the coloring of~$G$.
\end{algorithmic}
\end{algorithm}

Correctness of this recursive algorithm is immediate from the 
score-preserving nature of the reductions.

\begin{claim} \label{bc}
For a \csp instance $(G,S)$ where $G$ has $n$ nodes and $m$ edges,
and whose reduction tree per \algba has depth~$d$,
\algbc runs in time $O(n r^{3+d})$
and in linear space,~$O(L)$.
\end{claim}

Our main result follows immediately from 
Lemma~\ref{treeBlemma},
Lemma~\ref{depth19} 
(or Remark~\ref{depth4} for graphs with maximum degree 4 or less),
and Claims \ref{bb} and~\ref{bc}.

\begin{theorem} \label{maintheorem}
\algb solves a \mc instance $(G,S)$,
where $G$ has $n$ vertices and $m$ edges,
in time \timeb 
and in linear space,~$O(L)$.
If $G$ has maximum degree 4 the time bound may be replaced by
$O(\oneplus n r^{4+3m/16})$,
and if $G$ has maximum degree~3, by
$O(\oneplus n r^{3+m/6})$.
\end{theorem}

\section{Vertex-parametrized run time} \label{n-sec}

In most of this paper we consider run-time bounds as a function 
of the number of edges in a \mc instance's constraint graph, 
but we briefly present a couple of results giving time bounds
as a function of the number of \emph{vertices}, 
along with the average degree $\aved$ 
and (for comparison with existing results)
the maximum degree~$\Delta$.

For general \mc{s}, we derive a 
run-time bound
by using the following lemma in lieu of Lemma~\ref{depth19}.
(Thus, the linear-programming analysis plays no role here;
we are simply using the power of our reductions. 
Because the lemma bounds the \emph{number} of III-reductions,
not just their depth, it will also suffice to use \alga 
instead of the more complicated \algb.)

\begin{lemma} \label{decycle}
For a graph $G$ of order~$n$, with average degree~$\aved \geq 2$,
in time $\poly(n)$ we can find a reduction sequence with
at most $(1-\tfrac2{\aved+1}) n$ III-reductions.
\end{lemma}

\begin{proof}
Let $\a_2(G)$ be the maximum number of vertices in an induced forest in~$G$. 
This quantity was investigated by Alon, Kahn and Seymour \cite{decycle}, 
who showed that
$$\a_2(G)\ge\sum_{v\in V(G)}\min\left\{1,\frac{2}{d(v)+1}\right\},$$
and that there is a polynomial-time algorithm for finding 
an induced forest of the latter size 
(in fact, they proved a rather more general result; 
this is the special case of their Theorem~1.3 with 
degeneracy parameter~$2$).  
It follows (same special case of their Corollary~1.4) 
that if $G$ has average degree $\aved\ge2$ 
then $$\a_2(G)\ge\frac{2n}{\aved+1}.$$
Note that this is sharp when $G$ 
is a union of complete graphs of order $\aved+1$.

Now we simply III-reduce on every vertex of $G$ \emph{not} 
in the induced subgraph
(or 0-, I- or II-reduce on such a vertex which has degree $<3$
by the time we reduce on it).
After this sequence of reductions, the graph is a forest, 
and 0-, I- and II-reductions suffice to reduce it to the empty graph.
Thus the total number of III-reductions needed 
is $\leq n-\a_2(G) \leq n(1-\tfrac2{\aved+1})$.
\end{proof}

\begin{theorem} \label{n-param}
A \mc instance with constraint graph $G$ of order $n$ 
with average degree $\aved \ge 2$ can be solved in time
$$\O{n r^{\left(1-\frac{2}{\aved+1}\right)n} +\poly(n)} . $$
\end{theorem} 

\begin{proof}
Immediate from Lemma~\ref{decycle} and Claim~\ref{Aresult}.
(Since Lemma~\ref{decycle} gives a bound on the number of III-reductions,
not merely the depth, it suffices to use \alga rather than 
the more complicated \algb.)
\end{proof}

Note that for $\aved < (\sqrt{17561}+181)/38 \approx 8.25$, 
Theorem~\ref{maintheorem} gives a 
smaller bound than Theorem~\ref{n-param}, 
while for $\aved < 100/31 \approx 3.23$ 
the best bound is given by 
our $\Ostar{r^{(d-2)n/4}}$
algorithm from \cite{random03,linear}
(there stated more precisely as $\O{n r^{(m-n)/2}}$).

Theorem~\ref{n-param} improves upon one recent result of 
Della Croce, Kaminski, and Paschos~\cite{DCKP06},
which solves Max Cut (specifically) in time 
$\Ostar{2^{mn/(m+n)}} = \Ostar{2^{ (1-\frac2{d+2}) n}}$.
A second algorithm from \cite{DCKP06} solves Max Cut in time
$\Ostar{2^{ (1-2/\Delta) n}}$, 
where $\Delta$ is the constraint graph's maximum degree; 
this is better than our general algorithm if the
constraint graph is ``nearly regular'', with $\Delta<d+1$.

Our results also improve upon a recent result of 
F\"urer and Kasiviswanathan~\cite{FK07},
which, for binary \mc{s}, claims a running time of 
$\Ostarb{2^{(1-\tfrac{1}{\aved-1})n}}$
(when $\aved>2$ and the constraint graph is connected,
per personal communication).
For $\aved>3$ the bound of Theorem~\ref{n-param} is smaller, 
while for $2<\aved \leq 3$ (in fact, for $d$ up to~5), 
our $\Ostar{2^{n(d-2)/4}}$
algorithm from \cite{random03,linear} is best.

It is also possible to modify the algorithm described by 
Theorem~\ref{maintheorem} to give reasonably good 
vertex-parametrized algorithms for special cases,
such as Maximum Independent Set.
As remarked in the Introduction, however,
there are faster algorithms for MIS.

\begin{corollary} \label{MIS}
An instance of weighted Maximum Independent Set on an $n$-vertex graph
can be solved in time $\O{n2^{3n/8}}$
and in linear space, $O(m+n)$.
\end{corollary}

\begin{proof}
If $n < 20$ we may solve the instance in time~$O(1)$, 
and if the graph's maximum degree is $\Delta \leq 4$ 
we apply \algb, use Theorem~\ref{maintheorem}'s time bound of 
$O(\oneplus n r^{4+3m/16})$,
and observe that this is $O(\oneplus n 2^{3n/8})$.
Otherwise we use a very standard MIS reduction:
for any vertex~$v$,
either $v$ is not included in the independent set
or else it is and thus none of its neighbors is;
therefore the maximum weight of an independent set of $G$ satisfies 
$s(G) = \max\{s(G-v), \; w(v)+s(G-v-\Gamma(v))\}$, 
where $w(v)$ is the weight of vertex $v$ and $\Gamma(v)$ is its neighborhood.
``Rescaling'' time as usual so that we may drop the $O(\cdot)$ notation, 
if there is a vertex of degree 5 or more,
the running time $f(n)$ satisfies 
$f(n) \leq n+f(n-1)+f(n-6)$.
(A relevant constant for this recursion 
is $\alpha \colon 1=\alpha^{-1}+\alpha^{-6}$; 
its value is about $1.285$, and in particular less than $2^{3/8}$.)
For $n \geq 20$, 
induction on $n$ confirms that $f(n) \leq \oneplus n 2^{3n/8}$.
\end{proof}

\section{Treewidth and cubic graphs} \label{sec:treewidth}

In this section we show several connections between our LP method,
algorithms, and the treewidth of graphs, especially cubic (3-regular) graphs. 
We first define treewidth, 
and in Section~\ref{twbounds} show that it can be bounded in terms 
of our III-reduction depth. 
In Section~\ref{cubic} we show how a bound on the treewidth of cubic graphs 
can be incorporated into our LP method to give a treewidth bound 
for general graphs, and in turn faster (but exponential space) algorithms.
In Section~\ref{otheralgs} we show how fast algorithms for cubic graphs 
generally imply fast algorithms for general graphs, 
independent of treewidth. 

First, we recall the definition of treewidth and introduce
the notation we will use.
Where $G=(V,E)$, 
a \emph{tree decomposition} of $G$
is a pair $(X,T)$, 
where 
\begin{enumerate}
\item
$X=\{X_1,\dots,X_q\}$ is a collection of 
vertex subsets, called ``bags'', covering~$V$,
\ie, $X_i \subset V$ and $\bigcup_{i=1}^q X_i = V$;
\item
each edge of $G$ lies in some bag,
\ie, $(\forall uv \in E)(\exists i) \colon \{u,v\} \subset X_i$;
and
\item
$T$ is a tree on vertex set $X$ 
with the property that if $X_j$ lies on the path between $X_i$ and~$X_k$, 
then $X_j \supset (X_i \cap X_k)$.
\end{enumerate}
The width of the decomposition tree is defined as $\max_i |X_i|-1$, 
and a graph's treewidth is the minimum width over all tree decompositions.
Trees with at least one edge have treewidth~$1$,
and series-parallel graphs have treewidth at most~$2$.

 From Claim~\ref{bc} we have the following corollary. 

\begin{corollary} \label{series-parallel}
A CSP whose constraint graph $G$ is a tree or series-parallel graph
can be solved in time $O(r^3 n)$ and in linear space.
\end{corollary}

\begin{proof}
A tree $G$ can be reduced to a vertexless graph 
by 0- and I-reductions alone: it has III-reduction depth~0.
By definition, a series-parallel graph $G$
arises from repeated subdivision and duplication of a single edge.
It follows that II-reductions (with their fusings of multiple edges)
suffice to reduce $G$ to a collection of isolated edges 
(disjoint $K_2$'s), 
which are reduced to the vertexless graph by 0- and I-reductions.
Again, $G$ has III-reduction depth~0.
\end{proof}

\subsection{Implications of our results for treewidth} \label{twbounds}

Although trees and series-parallel graphs are both classes of graphs
with small treewidth and III-reduction depth~$0$, 
there is no reason to think that our algorithm will produce 
shallow III-reduction depth for all graphs of small treewidth. 
However, there is an implication in the opposite direction, 
per Claim~\ref{treewidth}.

\begin{lemma} \label{twreduce}
If $G$ is 0-reduced to a vertexless graph, $\tw(G)=0$.
If $G$ is I-reduced to $G'$, $\tw(G) = \max\set{1,\tw(G')}$.
If $G$ is II-reduced to $G'$, $\tw(G) = \tw(G')$.
If $G$ is III-reduced to components $G_1,\ldots,G_s$, 
$\tw(G) \leq 1+\max_i \tw(G_i)$.
\end{lemma}

\begin{proof}
For a 0-reduction, $G$ is a single vertex, which has treewidth~0.
Otherwise, first note that $\tw(G) \geq \tw(G')$, 
as shown by the tree decomposition for $G'$ 
induced by any tree decomposition of~$G$.

For a I-reduction,
$G$ adds a pendant edge $uv$ to some vertex $v$ of $G'$.
For any tree decomposition $(X',T')$ of $G'$, 
we can form a tree decomposition of $G$
by adding a new bag $X_0=\set{u,v}$ and 
linking it to any bag $X'_i \ni v$.
This satisfies the defining properties of a tree decomposition,
and has treewidth $\max\set{1,\tw(G')}$.

For a II-reduction, $G$ subdivides some edge $uv$ of $G'$ 
with a new vertex~$w$.
We mirror this in the decomposition tree in 
a way depending on two cases.
Either way, $(X',T')$ has a bag containing $\set{u,v}$.
If there is any bag of size 3 or more, 
we simply add a new bag $X_0 = \set{u,v,w}$
and link it to any bag $X'_i \supseteq \set{u,v}$.
If the maximum bag size is 2 then without loss of 
generality there is a single bag $X'_i= \set{u,v}$,
each of whose neighbors may contain either $u$ or $v$ but not both. 
We replace $X'_i$ with a pair of bags
$X_u=\set{u,w}$ and $X_v=\set{v,w}$, 
join them with an edge,
and join the former neighbors of $X'_i$ to either $X_u$ or $X_v$
depending on whether the neighbor contained $u$ or $v$
(if neither, the choice is arbitrary).
In either case 
this shows that $\tw(G) \leq \tw(G')$.

For a III-reduction on a vertex $v$,
let $(X^{(i)},T^{(i)})$ be tree decompositions of the components $G_i$
resulting from $v$'s deletion.
To obtain a tree decomposition of $G$,
first add $v$ to every bag of every tree;
every edge $(v,x)$ can be put in some such bag.
Also, create a new bag containing only the vertex $v$,
and join it to one (arbitrarily chosen) bag 
from each $(X^{(i)},T^{(i)})$,
thus creating a single tree and having the third defining property
of a tree decomposition.
This shows that $\tw(G) \leq 1+ \max_i \tw(G_i)$.
\end{proof}

\begin{claim} \label{treewidth}
If a graph $G$ has a reduction tree of III-reduction depth~$d$, 
then $G$ has treewidth~$\leq d+1$.
\end{claim}

\begin{proof}
 From the preceding lemma, the treewidth of $G$ 
is bounded by applying the various treewidth-reduction rules
along some critical (though typically not unique)
root-to-leaf path in the reduction tree. 
Traversing that path from leaf to root,
the case where treewidth changes from 0 to 1 
(from a I-reduction) occurs at most once, 
and otherwise the treewidth increases only at III-reduction nodes. 
Thus, $\tw(G)$ is at most 1 plus the maximum,
over all root-to-leaf paths, 
of the number of III-reductions in the path, 
which is to say $d+1$.
\end{proof}

\begin{corollary} \label{cor:treewidth}
A graph $G$ with $m$ edges has treewidth at most $3+19m/100$, 
and a tree decomposition of this width can be produced in time~$O(mn+n)$.
\end{corollary}

\begin{proof}
Immediate from the depth bound of Lemma~\ref{depth19}, 
\algba's running time per Lemma~\ref{treeBlemma}, 
and the algorithm in the proof of Claim~\ref{treewidth}.
\end{proof}

\subsection{Implications from treewidth of cubic graphs} \label{cubic}

In this section we explore how treewidth bounds for cubic graphs 
imply treewidth bounds for general graphs. 
Algorithmic implications of these treewidth bounds are discussed 
in the next subsection.

Building on a theorem of Monien and Preiss that
any cubic (3-regular) graph with $m$ edges has bisection width
at most $(1/9+\oo)m$ \cite{Monien},
Fomin and H{\o}ie show that such a graph
also has pathwidth at most $(1/9+\oo)m$ \cite{Fomin}.
(The $\oo$ terms here are as $m \rightarrow \infty$.)
For large $m$ 
this is significantly better than the treewidth bound of $1+m/6$ that would 
result from 
Claim~\ref{treewidth} and
the cubic III-reduction depth bound of $m/6$
(each III-reduction on a vertex of degree 3 destroying 6 edges).
Since we perform degree-3 III-reductions in a component only 
when it has no vertices of higher degree, 
it is possible to use this more efficient treatment of cubic graphs 
in place of our degree-3 III-reductions, 
as we now explain.

The result from \cite{Fomin} that a 3-regular graph with $m$ edges 
has pathwidth at most $(1/9+\oo)m$ implies the following lemma.
Since \cite{Fomin} relies on a polynomial-time construction, 
the lemma is also constructive.

\begin{lemma} \label{depth3reg}
If every 3-regular graph $G$ with $m$ edges has 
treewidth at most $\a m$, 
then any graph $G$ with $m$ edges has treewidth 
$\tw(G) \leq 3+\beta(\alpha) m$, 
and any graph of maximum degree $\D(G) \leq 4$ has
$\tw(G) \leq 2+\beta_4(\alpha) m$, 
where $\beta(\alpha)$ and $\b_4(\a)$ are given by Lemma~\ref{lpalpha}.
\end{lemma}

\begin{proof}
Recall that our graph reduction algorithm performed III-reductions
on vertices of degree 5 and 4 in preference to vertices of degree~3. 
Build the reduction tree as usual, 
but terminating at any node corresponding to a 
graph which is either vertexless \emph{or} 3-regular.
By Lemma~\ref{twreduce} and observations in the proof
of Lemma~\ref{treewidth},
the treewidth of the root (the original graph $G$)
is at most 1 plus the maximum, over all root-to-leaf paths, 
of the ``step count'' (or ``depth'') of each reduction
(1 for III-reductions, 0 for other reductions)
plus the treewidth of the leaf.
If we add a ``reduction'' taking an $m$-edge 3-regular graph
to a vertexless graph, and count it as $\a m$ steps, 
then $\tw(G)$ is at most 1 plus the maximum over all root-to-leaf paths
of the step counts along the path.

We may bound this value by the same LP approach taken previously.
We exclude the old degree-3 III-reduction,
characterized by line \arabic{line3} of Table~\ref{TableB}.
In its place we introduce a family of reductions:
for each number of edges $m'$ in a cubic graph (necessarily a multiple of~3) 
we have a reduction that counts as $\a m'$ steps and
destroys all $m'$ edges, all $2/3 m'$ degree-3 vertices of the cubic graph,
and 0 vertices of degrees 4 and~5.
As before, going down a path in the reduction tree, 
any ``bad'' reduction (a \badq or \badp reduction)
is either paired with a good one to make a combined reduction, 
or is counted as 0 steps (in at most 2 instances per path). 
The total number of reduction steps is thus at most 2 plus 
the step count of a feasible LP solution. 
Since a row of an LP may be rescaled without affecting the solution value, 
we may replace the family of 3-regular reductions 
with a single reduction that counts as $\a$ steps,
destroys $1$ edge and $2/3$ vertices of degree~3, 
and 0 vertices of degrees 4 and~5.
If this LP has optimal solution $\beta m$, 
then the path has true step count $\leq 2+\beta m$
and $G$ has treewidth $\tw(G) \leq 3+\beta m$.
The proof is completed by Lemma~\ref{lpalpha},
establishing $\beta$ as a function of~$\alpha$.
\end{proof}

\begin{lemma} \label{lpalpha}
Let LP be the linear program of Table~\ref{TableB}
whose line \arabic{line3} is replaced as below. 
\rm
\begin{center}
\begin{tabular}{|r@{\hspace*{0.4cm}}||c|ccccc||crcrr|r|c|}
\hline \padhline
\mbox{ }  & deg & \multicolumn{5}{c||}{\#nbrs of deg}
 & \multicolumn{5}{c|}{destroys} & forces & depth \\
\tabrownonum  \mbox{} 5 4 3 2 1  {$e$} 4 3 2 1 { } { } 
\hline \hline \padhline
\mbox{old} \tabrownonum 3  0  0  3  0  0  6  0  4      0  0   0   1
\mbox{new} \tabrownonum 3  0  0  3  0  0  1  0  {2/3}  0  0   0   {\alpha}
\hline
\end{tabular}
\end{center}
\it
Then LP has optimal solution 
\begin{align*}
\beta(\alpha) &= 
\begin{cases}
7/50+ (3/10) \a & 1/9 \leq \a \leq 1/5 
\\
13/75 & 0 \leq \a \leq 1/9 .
\end{cases}
\end{align*}
The same linear program restricted to the constraints
corresponding to reductions on vertices of degree 4 and smaller,
call it \LPfour, has optimal solution 
\begin{align*}
\beta_4(\alpha) &= 
\begin{cases}
1/8 + (3/8) \a & 1/9 \leq \a \leq 1/5 
\\
1/6 & 0 \leq \a \leq 1/9 .
\end{cases}
\end{align*}
\end{lemma}

\begin{proof}
To help give a feeling for the interpretation of our 
linear-programming analysis,
we will first give a very explicit duality-based proof, 
carrying it through for just one of the lemma's four cases. 
We will then show a much simpler proof method
and apply it to all the cases.

For the first case,
it suffices to produce feasible primal and dual LP solutions 
with the claimed costs. 
With $1/9 \leq \a \leq 1/5$, 
the primal solution puts weights exactly $0.30, 0.06, 0.08$ respectively
on the following rows of LP:
\begin{center}
\begin{tabular}{||c|ccccc||crrrr|r|c|}
\hline \padhline%
deg & \multicolumn{5}{c||}{\#nbrs of deg}
 & \multicolumn{5}{c|}{destroys} & forces & depth \\
\tabrownonumx  \mbox{} 5 4 3 2 1  {$e$} 4 3 2 1 { } { } 
\hline \hline 
\tabrownonumx 
\padhline
3  0  0  3  0  0  1  0  {2/3}  0  0   0   {\alpha}
\tabrownonumx
4  0  3  1  0  0  5  4 -2  0  0   0   1
\tabrownonumx
5  4  1  0  0  0  5 -3 -1  0  0   0   1
\hline
\end{tabular}
\end{center}
The solution is feasible because the weighted sum of the rows 
destroys exactly 1 edge and a nonnegative number 
(in fact,~0) of vertices of each degree.
The value of $\a$ does not enter into this at all:
$\a$ does not appear in the constraints, so
the primal solution is \emph{feasible} regardless of~$\a$.
The primal's \emph{value} is the dot product of $(0.30,0.06,0.08)$ with the 
``depth'' column $(\a,1,1)$, and matches the 
value of $\b$ claimed in the lemma. 

The dual solution is 
$\frac1{600} [(84,-18,-126,0,0,-60)+ \a(180,90,630,0,0,900)]$. 
It is dual-feasible because, interpreting these values as weights 
on (respectively) edges, vertices of degree 4, 3, 2, and~1, and forces,
for each row of LP the sum of the weights of edges and vertices destroyed,
and forces,
is at least the number of steps counted.
(The inequality is tight for the rows displayed above,
but one must check it for all rows.
For some rows, such as 
Table~\ref{TableB}'s line~11,
corresponding to reduction on 
a vertex of degree 5 with four neighbors of degree 4 and one of degree~5, 
the inequality is violated for $\a>1/5$.)
The dual LP value is the dot product of the dual solution with the 
primal's constraint vector $(1,0,0,0,0,0)$
(at least 1 edge, 0 vertices of each degree, 
and 0 ``forces'' should be destroyed).
Thus the dual value is
$1 \times (84+180 \a)/600 = 0.14 + 0.30 \alpha$,
matching the value specified in the lemma,
and thus also matching the primal value 
and proving the solution's optimality.

A much easier proof comes from exploiting a standard and simple fact 
from linear-programming sensitivity analysis:
Suppose a single vector $\xstar$ is an optimal solution to 
two linear programs with the same constraints but different objective
functions, given by vectors $c_1$ and $c_2$ respectively. 
Then $\xstar$ is also optimal for any linear program 
where again the constraints are the same, 
and the objective function $c$ is any
convex combination of $c_1$ and $c_2$.%
\footnote{%
Proof of this fact is instant: 
Optimality of $\xstar$ for $c_1$ means that for any feasible~$x$, 
$c_1 x \leq c_1 \xstar$, and likewise for $c_2$. 
Then for any convex combination $c=pc_1+qc_2$, $p+q=1$, 
$p,q \geq 0$, optimality of $\xstar$ for $c$ is proved by
the observation that for any feasible~$x$,
$cx= (pc_1+qc_2)x=p(c_1 x)+q(c_2 x) 
 \leq p(c_1 \xstar)+q(c_2 \xstar) = (pc_1+qc_2)\xstar = c \xstar$.%
}

Thus, to verify the case we have already done,
it suffices to check that a single primal solution $\xstar$
is optimal for both $\a=1/9$ and $\a=1/5$. 
This can easily be done by solving LP for some intermediate 
value, say $\a=1/7$, 
and checking that the primal $\xstar$ obtained, 
dotted with the objective vector corresponding to $\a=1/9$,
is equal to the solution value of the LP for $\a=1/9$,
and performing the same check for $\a=1/5$.
(Even easier, but not quite rigorous, is simply to solve the 
LP for, say, $\a = 1/9+0.001$ and $\a=1/5-0.001$, 
and verify that the two primal solutions are equal.)
The remaining cases are verified identically.
\end{proof}

\begin{corollary} \label{expcor}
Any graph $G$ with $m$ edges has $\tw(G) \leq (13/75+\oo)m$,
and if $\D(G)\leq 4$ then $\tw(G) \leq (1/6+\oo)m$.
\end{corollary}

\begin{proof}
Immediate from
Lemmas \ref{depth3reg} and \ref{lpalpha}, and
the fact that every cubic graph with $m$ edges 
has treewidth $\leq (1/9+\oo)m$~\cite{Fomin}. 
The additive constants can be absorbed into the~$\oo m$.
\end{proof}

We now discuss algorithmic implications of these treewidth bounds.

\subsection{Implications from algorithms for cubic graphs} 
 \label{otheralgs}

Efficient algorithms for constraint satisfaction of various sorts,
and related problems, on graphs of small treewidth
have been studied since at least the mid-1980s,
with systematic approaches dating back at least 
to \cite{Dechter87,Dechter89,Arnborg}. 
A special issue of Discrete Applied Mathematics
was devoted to this and related topics in 1994 \cite{DAM},
and the field remains an extremely active area of research.

It is something of a folk theorem that a \mc instance of treewidth
$k$ can be solved in time \emph{and space} $\Ostar{r^k}$ 
through dynamic programming.
(The need for exponential space is of course a 
serious practical drawback.)
Such a procedure was detailed by 
Jansen, Karpinski, Lingas and Seidel \cite{Karpinski}
for solving maximum bisection, minimum bisection, and maximum clique. 
Those problems are in fact slightly outside the \mc framework
defined here, but within a broader framework of 
``\Generalized CSPs'' that we explore in~\cite{CountingArxiv,CountingArxiv2}.
In \cite{CountingArxiv,CountingArxiv2} 
we show how to use dynamic programming on 
tree decompositions of width $k$
to solve any Polynomial CSP, 
including the problems above and any \mc, in time and space 
$\Ostar{r^k}$.

Direct application of dynamic programming in conjunction with 
Corollary~\ref{expcor} 
means that
any \mc can be solved in 
time and space~$\Ostar{r^{(13/75+\oo)m}}$.
However, we can do better.

Similarly to how 
Lemma~\ref{depth3reg} showed that a cubic treewidth bound $\a m$ 
implies a general treewidth bound of $\b m$, 
Theorem~\ref{cubicf} shows that
an $\Ostar{r^{\a m}}$-time algorithm for cubic instances of \mc
can be used to construct an $\Ostar{r^{\b m}}$-time algorithm 
for arbitrary instances.
The approach gives greater generality, 
since the algorithm for cubic instances need not have 
anything to do with treewidth.
And when the algorithm for cubic instances \emph{is} 
tree decomposition-based dynamic programming,
this approach gives greater efficiency:
we can match the previous paragraph's time bound,
while reducing the space requirement (Corollary~\ref{generic}).

\begin{theorem} \label{cubicf}
Given a value $\a>0$, an integer~$r$, 
and a function $g(m) = \Ostar{r^{\a m}}$,
suppose there is an algorithm that,
for any $m$-edge 3-regular graph~$G$, 
solves any CSP with constraint graph $G$ and domain $[r]$
in time~$g(m)$. 
Then there is an algorithm which solves any CSP with domain $[r]$
and any $m$-edge constraint graph $G$ in time $\Ostarr r {\b(\a) m}$,
and in time $\Ostarr r {\b_4(\a) m}$ if $\D(G)\leq 4$,
with $\b(\a)$ and $\b_4(\a)$ given by Lemma~\ref{lpalpha}.
If the hypothesized algorithm is guaranteed to solve an instance of 
input size $L$
using space $O(s(L))$, for some nondecreasing function~$s$,
then the algorithm assured by the theorem uses space $O(L+s(L))$.
\end{theorem}

\begin{proof}
The proof is similar to that of Lemma~\ref{depth3reg}. 
Introduce a family of ``reductions'' reducing an
$m$-edge 3-regular graph to
a vertexless graph and counting for depth $\a m$. 
Precisely as in the earlier proof,
represent them all in the LP by a single reduction
destroying 1 edge, $2/3$ vertices of degree~3, 
and 0 vertices of degree 4 and 5, 
and counting as depth~$\a$. 

Reduce a graph $G$ as far as possible by 0-, I- and II-reductions, and
III-reductions on vertices of degree 4 and above.
For any tree node, and corresponding reduced constraint graph $G'$,
define the depth of $G'$ to be the maximum, 
over all its 3-regular 
leaf instances $G_i$ having $m_i$ edges respectively,
of $\a m_i$ plus the number of III-reductions to get from $G$ to~$G_i$. 
 From our usual LP setup and Lemma~\ref{lpalpha}, it is immediate that 
any $m'$-edge graph $G'$ has depth $\leq 2+\b m'$. 

It remains only to show that depth $2+\b m$ implies
running time $\Ostarr r {2+\b m}$, and we will do this inductively. 
Note that a cubic graph with $m$ edges has $n=2m/3$ vertices,
so the fact that $g(m) = \Ostarr r {\b m}$
implies that there is some polynomial $p(n)$ 
such that $g(m) \leq p(n) f(n,\b m)$, 
where $f$ is the function defined by \eqref{recurbound}
in the proof of Claim~\ref{bb}.
Without loss of generality, assume $p(n) \geq 1$.
Note that 
$f$ is given explicitly, and $p$ depends on the 
bound $g$ guaranteed by the Theorem's hypothesis, 
but not on $r$, 
$G$, etc.

Suppose the original instance's constraint graph $G$ has $n$ vertices. 
We now show inductively that
each reduced instance $G'$ 
with $n'$ vertices and depth $d'$ can be solved in 
time $p(n) f(n',d')$.
(We really do mean $p(n)$, not $p(n')$.)
The induction begins at the leaves, and proceeds up the tree.
For a leaf $G'$, which is a 3-regular instance, 
the property is guaranteed 
by the theorem's hypothesis, $d'= \b m'$, and $p(n)\geq 1$. 
Otherwise, for a node $G'$ we may inductively assume the property holds for
its children, in which case the running time for $G'$ is at most
\begin{align*}
 r^3 n' + r \sum p(n) f(n_i, d_i)
 & \leq p(n) [r^3 n' + r \sum f(n_i, d-1)
 \\ & \leq p(n) f(n', d) ,
\end{align*}
where the second inequality is precisely the calculation 
performed after \eqref{recurbound}.
Taking $G'=G$ shows that the root node $G$ can be solved in time
$\leq p(n) f(n, \b m)$ $=\Ostarr r {\b m}$.

Except for the calls on the hypothesized algorithm, 
our overall algorithm uses space $O(L)$, 
per Theorem~\ref{thmA}. 
Since each cubic subinstance has size at most $L$, and $s$ is nondecreasing, 
the total space needed is $O(L+s(L))$.
\end{proof}

\begin{corollary} \label{generic}
A \mc instance with domain size $r$ and $m$ dyadic constraints can be solved 
in time $\Ostar{r^{(13/75+\oo)m}}$,
and if $\D(G) \leq 4$, time $\Ostar{r^{(1/6+\oo)m}}$, 
in either case in space $\Ostar{r^{(1/9+\oo)m}}$.
\end{corollary}

\begin{proof}
With $\a=1/9+\oo$, Theorem~\ref{cubicf}'s hypothesized algorithm 
for $m$-edge cubic instances 
is given by dynamic programming on a tree decomposition 
of treewidth $\leq \a m$
(which by \cite{Fomin} exists 
and can be found in polynomial space and time),
and runs in space and time $\Ostar{r^{\a m}}$.
The Corollary follows from Theorem~\ref{cubicf}.
\end{proof}

\begin{remark} \label{stuck}
While it would be nice to reduce the treewidth bound of a 
cubic graph from the $(1/9+\oo)m$ of \cite{Fomin}
to a simple~$m/9$,
any further reduction (e.g., to $m/10$) would result in 
{no improvement} in Corollaries \ref{expcor} or~\ref{generic},
unless accompanied by improvements in some other aspect of the analysis.
\end{remark}

While surprising, this fact is instantly obvious 
from the linear-programming results of Lemma~\ref{lpalpha}.
One interpretation is that it happens because, for $\a<1/9$,
the primal solution has weight 0 on the degree-3 III-reduction.

\section{Conclusions} \label{conclusions}
As noted in the Introduction,
linear programming 
is key to our algorithm design
as well as the analysis. 
We begin with a collection of reductions, and a preference order 
on them, guided by intuition.
The preference order both excludes some cases
(e.g., reducing on high-degree vertices first, 
we do not need to worry about a reduction vertex having 
a neighbor of larger degree)
and determines an LP. 
Solving the LP pinpoints the ``bad'' reductions that determine the bound.
We then try to ameliorate these cases: 
in the present paper we showed that each could be paired with another
reduction to give a less bad combined reduction, 
but we might also have taken some other course such as 
changing the preference order to eliminate bad reductions.
Using the LP as a black box is a convenient way to engage in this 
cycle of algorithm analysis and improvement,
an approach that should be applicable to other problems. 

While we focus on the linear program as a way to bound
our key parameter, a graph's III-reduction depth, 
Section~\ref{sec:treewidth} shows that it also applies to treewidth.
Sharper results for (constraint) graphs of maximum degree 4
can be obtained simply by pruning down the LP. 

Because the LP's dual solution can be interpreted as a set of 
weights on edges and vertices of various degrees, 
the LP method introduced in \cite{random03}, and further developed here,
is closely connected to a potential-function approach. 
The determination of optimal weights can always be expressed as 
an optimization problem 
(see Eppstein's~\cite{EppsteinQuasi} and the survey~\cite{FominSurvey}), but
its expression as an LP seems limited to cases where the CSP reductions 
are ``symmetric'' in the sense that 
they yield a single reduced graph.
(A natural independent-set reduction is not symmetric in this sense,
as reducing on a vertex $v$ 
yields two reduced instances with different graphs:
one deleting only the vertex $v$, 
the other also deleting all $v$'s neighbors.)
However, it can still be possible to plug bounds derived from 
asymmetric reductions into the LP method;
for example the hypothesized algorithm in Theorem~\ref{cubicf}
might depend on asymmetric reductions.
When the LP method is applicable, provably optimal weights 
are efficiently obtainable. 
Linear programming also provides an elegant framework
and points the way to structural results like Lemma~\ref{lpalpha},
but similar results could also be obtained under weaker conditions,
outside the LP framework.
For example, to prove Lemma~\ref{lpalpha}, 
convexity of the solution space and linearity 
of the objective function would have sufficed.

It must be emphasized that the improvement of the
present $19m/100$ depth bound over the previous $m/5$ 
is not a matter of a more detailed case analysis;
indeed there are far fewer cases here than in most 
reduction-based CSP algorithms.
Ultimately, the improvement comes from exploiting the 
constraint graph's division into components. 
While this is very natural, its use in
combination with the reduction approach and LP analysis 
is slightly tricky, and appears to be novel.

Linear programming aside, our approach 
seems not to extend to 3-variable CSPs, 
since a II-reduction would combine two 3-variable clauses into 
a 4-variable clause.

The improvement from $m/5$ to $19m/100$ is significant in that $m/6$
appears to be a natural barrier:
In a random cubic graph, a III-reduction results in 
the deletion of 6 edges and a new cubic graph, 
and to beat $m/6$ requires either distinguishing the new graph
from random cubic, or targeting a set of III-reductions to divide 
the graph into components. 
Such an approach would require new ideas outside the scope of the
local properties we consider here.

Finally, we remark that it would be interesting to analyze further the
behavior of algorithms on random instances.  
For example, it is shown in \cite{linear} that for any $c\le1$, 
\emph{any} \mc instance with constraint graph $G\in{\mathcal G}(n,p)$ can be solved in 
\emph{linear expected time}.
(Note that this is much stronger than succeeding in linear time
\emph{with high probability}.)  
Could this be extended to other problems?  
Could $2^{o(n)}$ runtime bounds be proved for random instances 
of problems such as Max Cut and Max 2-Sat with $cn$ clauses, 
where $c\gg 1$?  
What about approximation results?

\section*{Acknowledgments} 
We thank an anonymous referee and Daniel Raible for helpful comments.
The first author's research was supported in part by EPSRC 
grant GR/S26323/01. 
We are also grateful to the LMS for supporting two research visits.

\providecommand{\bysame}{\leavevmode\hbox to3em{\hrulefill}\thinspace}
\providecommand{\MR}{\relax\ifhmode\unskip\space\fi MR }
\providecommand{\MRhref}[2]{%
  \href{http://www.ams.org/mathscinet-getitem?mr=#1}{#2}
}
\providecommand{\href}[2]{#2}


\begin{thebibliography}{KMRR05}

\bibitem[AHe94]{DAM}
S.~Arnborg, S.T. Hedetniemi, and A.~Proskurowski (editors), \emph{Special issue
  on efficient algorithms and partial $k$-trees}, Discrete Applied Mathematics
  \textbf{54} (1994), no.~2-3.

\bibitem[AKS87]{decycle}
N.~Alon, J.~Kahn, and P.~D. Seymour, \emph{Large induced degenerate subgraphs},
  Graphs Combin. \textbf{3} (1987), no.~3, 203--211. \MR{MR903609 (88i:05104)}

\bibitem[AP89]{Arnborg}
S.~Arnborg and A.~Proskurowski, \emph{Linear time algorithms for {NP}-hard
  problems restricted to partial $k$-trees}, Discrete Applied Mathematics
  \textbf{23} (1989), no.~1, 11--24.

\bibitem[BE05]{Epps}
Richard Beigel and David Eppstein, \emph{3-coloring in time ${O}(1.3289^n)$},
  J. Algorithms \textbf{54} (2005), no.~2, 168--204.

\bibitem[Cre95]{creignou}
Nadia Creignou, \emph{A dichotomy theorem for maximum generalized
  satisfiability problems}, J. Comput. System Sci. \textbf{51} (1995), no.~3,
  511--522, 24th Annual ACM Symposium on the Theory of Computing (Victoria, BC,
  1992). \MR{MR1368916 (97a:68076)}

\bibitem[DCKP]{DCKP06}
F.~Della~Croce, Marcin~J. Kaminski, and Vangelis~Th. Paschos, \emph{An exact
  algorithm for {MAX-CUT} in sparse graphs}, Operations Research Letters, to
  appear (available online 2006, doi:10.1016/j.orl.2006.04.001).

\bibitem[DJ02]{Dahllof2002}
Vilhelm Dahll\"of and Peter Jonsson, \emph{An algorithm for counting maximum
  weighted independent sets and its applications}, Proceedings of the 13th
  Annual ACM-SIAM Symposium on Discrete Algorithms (SODA), January 2002.

\bibitem[DJW05]{Jonsson05}
Vilhelm Dahll{\"o}f, Peter Jonsson, and Magnus Wahlstr{\"o}m, \emph{Counting
  models for 2{SAT} and 3{SAT} formulae}, Theoret. Comput. Sci. \textbf{332}
  (2005), no.~1-3, 265--291. \MR{MR2122506 (2005j:68055)}

\bibitem[DP87]{Dechter87}
R.~Dechter and J.~Pearl, \emph{Network-based heuristics for
  constraint-satisfaction problems}, Artif. Intell. \textbf{34} (1987), no.~1,
  1--38.

\bibitem[DP89]{Dechter89}
Rina Dechter and Judea Pearl, \emph{Tree clustering for constraint networks
  (research note)}, Artif. Intell. \textbf{38} (1989), no.~3, 353--366.

\bibitem[Epp04]{EppsteinQuasi}
David Eppstein, \emph{{Quasiconvex analysis of backtracking algorithms}},
  Proceedings of the 15th Annual {ACM--SIAM} Symposium on Discrete Algorithms
  (New Orleans, LA, 2004) (New York), ACM, 2004, pp.~781--790.

\bibitem[FGK05]{FominSurvey}
Fedor~V. Fomin, Fabrizio Grandoni, and Dieter Kratsch, \emph{Some new
  techniques in design and analysis of exact (exponential) algorithms},
  September 2005.

\bibitem[FH06]{Fomin}
Fedor~V. Fomin and Kjartan H{\o}ie, \emph{Pathwidth of cubic graphs and exact
  algorithms}, Inform. Process. Lett. \textbf{97} (2006), no.~5, 191--196.
  \MR{MR2195217 (2006g:05199)}

\bibitem[FK05]{Furer05}
Martin F\"urer and Shiva~Prasad Kasiviswanathan, \emph{Algorithms for counting
  {2-SAT} solutions and colorings with applications}, Tech. Report TR05-033,
  Electronic Colloquium on Computational Complexity, March 2005, See {\tt
  http://www\allowbreak .eccc\allowbreak .uni-trier\allowbreak .de/eccc/}.

\bibitem[FK07]{FK07}
Martin Furer and Shiva~P. Kasiviswanathan, \emph{Exact {M}ax 2-{SAT}: Easier
  and faster}, Proceedings of SOFSEM 2007, LNCS 4362, Springer, 2007.

\bibitem[GHNR03]{Gramm03}
Jens Gramm, Edward~A. Hirsch, Rolf Niedermeier, and Peter Rossmanith,
  \emph{Worst-case upper bounds for {MAX}-2-{SAT} with an application to
  {MAX}-{CUT}}, Discrete Appl. Math. \textbf{130} (2003), no.~2, 139--155.
  \MR{MR2014655 (2004j:68077)}

\bibitem[Hir00]{Hirsch}
Edward~A. Hirsch, \emph{A new algorithm for {MAX}-2-{SAT}}, STACS 2000 (Lille),
  Lecture Notes in Comput. Sci., vol. 1770, Springer, Berlin, 2000, pp.~65--73.

\bibitem[JKLS05]{Karpinski}
Klaus Jansen, Marek Karpinski, Andrzej Lingas, and Eike Seidel,
  \emph{Polynomial time approximation schemes for max-bisection on planar and
  geometric graphs}, SIAM J. Comput. \textbf{35} (2005), no.~1, 110--119
  (electronic). \MR{MR2178800 (2006f:68143)}

\bibitem[KF02]{fedin}
Alexander~S. Kulikov and Sergey~S. Fedin, \emph{Solution of the maximum cut
  problem in time {$2^{\vert E\vert /4}$}}, Zap. Nauchn. Sem. S.-Peterburg.
  Otdel. Mat. Inst. Steklov. (POMI) \textbf{293} (2002), no.~Teor. Slozhn.
  Vychisl. 7, 129--138, 183.

\bibitem[KK06]{KK06}
A.~Kojevnikov and A.~S. Kulikov, \emph{A new approach to proving upper bounds
  for {MAX-2-SAT}}, Proceedings of the 17th Annual {ACM--SIAM} Symposium on
  Discrete Algorithms (Miami, FL, 2006) (New York), ACM, 2006, pp.~11--17.

\bibitem[KMRR05]{Kneis4}
Joachim Kneis, Daniel M{\"o}lle, Stefan Richter, and Peter Rossmanith,
  \emph{Algorithms based on the treewidth of sparse graphs}, Graph-theoretic
  concepts in computer science, Lecture Notes in Comput. Sci., vol. 3787,
  Springer, Berlin, 2005, pp.~385--396. \MR{MR2213887}

\bibitem[KR05]{Kneis05}
Joachim Kneis and Peter Rossmanith, \emph{A new satisfiability algorithm with
  applications to {M}ax-{C}ut}, Tech. Report AIB-2005-08, Department of
  Computer Science, RWTH Aachen, 2005.

\bibitem[KSTW01]{khanna}
Sanjeev Khanna, Madhu Sudan, Luca Trevisan, and David~P. Williamson, \emph{The
  approximability of constraint satisfaction problems}, SIAM J. Comput.
  \textbf{30} (2001), no.~6, 1863--1920 (electronic). \MR{MR1856561
  (2002k:68058)}

\bibitem[Mar04]{marx}
D\'aniel Marx, \emph{Parameterized complexity of constraint satisfaction
  problems}, Proceedings of the 19th {IEEE} Annual Conference on Computational
  Complexity {(CCC'04)}, 2004, pp.~139--149.

\bibitem[MP06]{Monien}
Burkhard Monien and Robert Preis, \emph{Upper bounds on the bisection width of
  3- and 4-regular graphs}, J. Discrete Algorithms \textbf{4} (2006), no.~3,
  475--498. \MR{MR2258338}

\bibitem[NR00]{NiRo00}
Rolf Niedermeier and Peter Rossmanith, \emph{New upper bounds for maximum
  satisfiability}, J. Algorithms \textbf{36} (2000), no.~1, 63--88.

\bibitem[Sch99]{schoening}
Uwe Sch{\"o}ning, \emph{A probabilistic algorithm for {$k$}-{SAT} and
  constraint satisfaction problems}, 40th Annual Symposium on Foundations of
  Computer Science (New York, 1999), IEEE Computer Soc., Los Alamitos, CA,
  1999, pp.~410--414. \MR{MR1917579}

\bibitem[SS03]{random03}
Alexander~D. Scott and Gregory~B. Sorkin, \emph{Faster algorithms for {MAX CUT}
  and {MAX CSP}, with polynomial expected time for sparse instances}, Proc. 7th
  International Workshop on Randomization and Approximation Techniques in
  Computer Science, RANDOM 2003, Lecture Notes in Comput. Sci., vol. 2764,
  Springer, August 2003, pp.~382--395.

\bibitem[SS04]{FasterIBM}
\bysame, \emph{A faster exponential-time algorithm for {Max 2-{S}at, {M}ax
  {C}ut, and {M}ax $k$-{C}ut}}, Tech. Report RC23456 (W0412-001), IBM Research
  Report, December 2004, See {http://\allowbreak domino\allowbreak
  .research\allowbreak .ibm\allowbreak .com/\allowbreak library/\allowbreak
  cyberdig.nsf}.

\bibitem[SS06a]{CountingArxiv}
\bysame, \emph{Generalized constraint satisfaction problems}, Tech. Report
  cs:DM/0604079v1, arxiv.org, April 2006, See {http://\allowbreak
  arxiv.org/\allowbreak abs/\allowbreak cs.DM/\allowbreak 0604079}.

\bibitem[SS06b]{FasterESA}
\bysame, \emph{An {LP}-designed algorithm for constraint satisfaction}, Proc.
  14th Annual European Symposium on Algorithms, {ESA}, ({Z\"urich},
  Switzerland, 2006), Lecture Notes in Comput. Sci., vol. 4168, Springer,
  September 2006, pp.~588--599.

\bibitem[SS06c]{linear}
\bysame, \emph{Solving sparse random instances of {Max Cut} and {Max 2-CSP} in
  linear expected time}, Comb. Probab. Comput. \textbf{15} (2006), no.~1-2,
  281--315.

\bibitem[SS07]{CountingArxiv2}
\bysame, \emph{Polynomial constraint satisfaction: A framework for counting and
  sampling {CSP}s and other problems}, Tech. Report cs:DM/0604079v2, arxiv.org,
  February 2007, See {http://\allowbreak arxiv.org/\allowbreak abs/\allowbreak
  cs.DM/\allowbreak 0604079}.

\bibitem[TSSW00]{SSTW00}
Luca Trevisan, Gregory~B. Sorkin, Madhu Sudan, and David~P. Williamson,
  \emph{Gadgets, approximation, and linear programming}, SIAM J. Comput.
  \textbf{29} (2000), no.~6, 2074--2097.

\bibitem[Wil04]{Williams}
Ryan Williams, \emph{A new algorithm for optimal constraint satisfaction and
  its implications}, Proc.\ 31st International Colloquium on Automata,
  Languages and Programming (ICALP), 2004.

\end{thebibliography}
\end{document}